\documentclass[]{aastex631}

\usepackage{amsmath}

\usepackage [english]{babel}
\usepackage [autostyle, english = american]{csquotes}
\MakeOuterQuote{"}

\shorttitle{Survey of Model Fits Through the L-T Sequence}
\shortauthors{Turner et al.}

\graphicspath{{./}{figures/}}

\begin{document}

\title{A Survey Of Model Fits to Brown Dwarf Spectra Through the L-T Sequence}

\correspondingauthor{Savanah K. Turner}
\email{savanahkayturner@gmail.com}

\author[0009-0009-9307-7237]{Savanah K. Turner}
\affiliation{Department of Physics and Astronomy\\ N284 ESC, Brigham Young University\\Provo, UT, 84602, USA}

\author[0000-0003-4658-7567]{Denise C. Stephens}
\affiliation{Department of Physics and Astronomy\\ N284 ESC, Brigham Young University\\Provo, UT, 84602, USA}
\email{denise\_stephens@byu.edu}

\author{Conner B. Scoresby}
\affiliation{Apache Point Observatory\\ 2001 Apache Point Road P.O. Box 59\\Sunspot, NM 88349, USA}
\email{cscoresby@apo.nmsu.edu}

\author[0000-0002-9192-0317]{Josh A. Miller}
\affiliation{Department of Physics and Astronomy\\ N284 ESC, Brigham Young University\\Provo, UT, 84602, USA}
\email{jam469@byu.edu}

\begin{abstract}

We fit archival near-infrared spectra of $\sim$300 brown dwarfs with atmosphere models from the Sonora and Phoenix groups. Using the parameters of the best-fit models as estimates for the physical properties of the brown dwarfs in our sample, we have performed a survey of how brown dwarf atmospheres evolve with spectral type and temperature. We present the fit results and observed trends. {We find that clouds have a more significant impact on near infrared spectra than disequilibrium chemistry, and that silicate clouds influence the near infrared spectrum through the late T types. }We note where current atmosphere models are able to replicate the data and where the models and data conflict. We also categorize objects with {similar spectral morphologies} into families and discuss possible causes for their unique spectral traits. {We identify two spectral families with morphologies that are likely indicative of binarity.}

\end{abstract}


\section{Introduction} \label{sec:intro}

The advent of powerful infrared telescopes such as the James Webb Space Telescope (JWST) has ushered in a new era of exoplanet discovery and characterization. The search for habitable worlds holds the interest of the science community and public alike. However, detailed spectroscopic study of an exoplanet's atmosphere is difficult due to the brightness of the planet's host star. 

Brown dwarfs are similar in size, temperature, and composition to hot planets, but can form without a host star. These "super Jupiters" \citep{superJupiter} can thus serve as more easily-accessible laboratories for studying the physics and chemistry of planetary atmospheres.

In this project, we seek to understand how the atmospheric properties of a brown dwarf change as it cools through the L-T spectral sequence. Complex cloud physics can be found in the atmospheres of brown dwarfs through this sequence, particularly as a brown dwarf cools through the L/T transition. Improving our understanding of the characteristics of L- and T- type atmospheres and how these characteristics evolve with spectral type can help us to better understand the properties of hot exoplanetary atmospheres.

\subsection{The L-T Spectral Type Sequence} \label{subsec:LTTypes}

L-type brown dwarfs range in temperature from about 2400 K to 1300 K. Their spectra are dominated by metallic hydrides and neutral alkali metals in the far optical and CO in the near infrared \citep{Kirkpatrick_1999}. {Their atmospheres have cloud layers of refractory materials such as} corundum, liquid iron, and silicate minerals such as enstatite, forsterite, and quartz \citep{Ackerman_2001,Allard_2001,Lodders_2004,Cushing_2008,Burningham_2021}. 

The effective temperatures of T-type brown dwarfs run from about 1300 K to 700 K. Unlike L-type objects, T-type spectra show CH$_4$ absorption features at 1.6 and 2.2 $\mu$m that become more pronounced for each progressive T subtype. The water band near 0.93 $\mu$m also deepens through the T sequence \citep{burgasser2001classification}. The silicate and iron cloud features of L-types are not seen in T-types, but laboratory condensation curves predict that T-type dwarfs could have thin salt clouds \citep{Morley_2012,Marley_2010}.

The initial temperature/spectral type of a newly-formed brown dwarf is determined by its mass. Because these "failed stars" are unable to sustain hydrogen fusion, after a period of deuterium fusion a brown dwarf spends the rest of its life cooling over time via blackbody radiation. As it cools, the object evolves through later spectral types. Thus, brown dwarf spectral types represent an evolutionary sequence rather than a one-time classification. 

As a brown dwarf cools through the spectral type sequence, it experiences interesting shifts in cloud properties, molecular abundances, and surface gravity (due to ongoing contraction). The transition region between the L and T spectral types is particularly interesting. With the decrease in temperature, the iron and silicate clouds that dominated the L-dwarf spectra condense deeper in the atmosphere and holes between the cloud layers develop, as is evident from the photometric variability observed in many late-L and early-T dwarfs \citep{Radigan_2014}. These holes allow light from deeper in the atmosphere to escape \citep{Ackerman_2001,Burgasser_2002}. At the same time, the dominant carbon-bearing molecule transitions from CO to CH$_4$.

Recent studies have shown that vertical mixing also plays an important role in brown dwarf atmospheres. Observations at 4.6 $\mu$m have shown that CO remains abundant in the atmosphere longer than expected through the L/T transition.  This suggests that vertical mixing is dredging CO up from deeper in the atmosphere, reducing the abundance of CH$_4$ higher in the atmosphere and driving the system out of equilibrium \citep{Sorahana_2012, Noll_1997,Miles_2020}. NH$_3$ and N$_2$ experience similar effects \citep{Cushing_2008}.

\subsection{Model Fit Surveys}\label{subsec:Surveys}

Much of what we know about the composition of planetary atmospheres has been learned by comparing observations to theoretical atmosphere models. When the theoretical models fit the data well, they can be used to turn a spectrum into a story that tells us about the real physical characteristics of the target object's atmosphere. Areas where models do not fit the data well can be just as enlightening; they show us the gaps in our understanding of atmospheric physics and chemistry and drive us to improve our theory.

The literature shows that performing model fits to a sample of brown dwarfs with a range of spectral types can provide important insights into how atmospheres evolve as they cool and also highlight the strengths/weaknesses of the model set. 

For example, {\citet{cushing06}} visually compared Spitzer IRS spectra of 46 brown dwarfs with spectral types M, L, or T to models from \citet{Allard_2001}, \citet{Marley_2002}, and \citet{Saumon_2008}. Their survey showed that the models did not completely fit the silicate features of mid L-dwarfs around 10 $\mu$m. They suggested that future models should incorporate a larger range of grain sizes {and more options for specific silicate species and crystallinity.}

\citet{Stephens_2009} fit models from \citet{Saumon_2008} to 1-15 $\mu$m spectra of 14 objects with spectral types L3.5-T5.5 and found trends in how vertical mixing and cloud properties change through the L/T transition. They also were able to highlight which spectral regimes are most impacted by changes in temperature, log($g$), vertical mixing, and cloud sedimentation efficiency $f_{sed}$.

\citet{Lueber_2022} used the open-source retrieval code HELIOS-r2 to fit near-infrared SpeX Prism spectra of 19 objects spanning the L and T types. They found that models with different cloud properties yielded different best-fit values of log($g$) for the same object, suggesting that low-resolution near-infrared data alone might not be sufficient to disentangle the effects of log($g$) and clouds.

\citet{Su_rez_2023} fit archival and new Spitzer IRS data to models and found that the silicate clouds of older, bluer mid L-dwarfs generally have smaller grains than the silicate clouds of young, red mid L-dwarfs. 

{\citet{zhang2021b} used the Bayesian framework Starfish \citep{Czekala_2015} and the cloudless, chemical equilibrium Sonora Bobcat models \citep{Marley_2021} to perform careful forward-model fits to 1-2.5 $\mu$m spectra of 55 brown dwarfs with spectral types T7-T9. Their work found unphysically small masses, unlikely young ages, and suprisingly low metallicities for the objects in their sample, leading them to conclude that clouds and/or disequilibrium chemistry are necessary to properly model late T-dwarfs. In a follow-up paper \cite{zhang2024} used the same Starfish framework to fit BT-Settl \citep{Allard_2012,2015A&A...577A..42B} forward models to 90 late-M and L dwarfs in young moving groups. Model-data mismatches in the peak \textit{J}- and \textit{H}-band fluxes resulted in their conclusion that a higher dust content needed to be added to the models in order to fit objects in this spectral range.}

These examples, along with others in the literature, demonstrate how model-fitting surveys can lead both to revised understanding of atmospheric processes and improved modeling techniques. 

\subsection{Testing Forward Model Fits}\label{subsec:Testing}

Many of the conclusions listed in Section \ref{subsec:Surveys} rely on the assumption that the best-fit model parameters (i.e. temperature, surface gravity, metallicity, etc.) are truly representative of the physical characteristics of the sample. Before forward model fits can be used to draw such conclusions, it is critical that the fit results be tested and benchmarked.

For example, \cite{zhang2021a} use benchmark brown dwarfs with empirically-determined characteristics and evolutionary models, which are generally considered to be more robust than atmospheric forward models, to test the accuracy of their forward-modeling framework and the atmosphere models on which it relies. They begin by identifying three brown dwarfs in systems with primary stars that have empirically-known distances, metallicities, and ages. They then use the Sonora Bobcat evolutionary models \citep{Marley_2021} and the known primary star ages and metallicities to derive the $T_{eff}$, log($g$), metallicities, radii, masses, and bolometric luminosities of the benchmark objects. Finally, they fit the Sonora Bobcat atmosphere models to near-infrared spectra of the benchmark objects. The authors compare the parameters derived from the evolutionary models to those obtained by the forward model fits, observing that either T$_{eff}$ and log($g$) or radius and metallicity are consistent between the two methods, but never all four parameters. They conclude that the discrepancy arises from imperfections in the atmospheric models, and that the two parameters that are consistent for each object depend on its T$_{eff}$. 

Similarly, \cite{sanghi2023hawaii} calculate bolometric luminosities and estimate ages for the objects in their sample. They then use the evolutionary models from \cite{Saumon_2008} and \cite{2015A&A...577A..42B} to derive masses, $T_{eff}$, log($g$), and radii. They compare their overall trends in evolutionary-model-derived parameters vs spectral type to corresponding trends in atmospheric-model-derived parameters vs spectral type. These authors observe that their evolutionary and atmospheric parameters differ the most significantly around the M/L spectral type transition.

Both \cite{zhang2021a} and \cite{sanghi2023hawaii} note that differences in atmospheric vs evolutionary parameters are not consistent across their sample, but rather vary with T$_{eff}$/spectral type. One model set might give accurate values for physical parameters for some spectral types but incorrectly constrain the same parameters for later/earlier types. Thus, it is critical that model fitting surveys test results across the full spectral type range of their sample.

\subsection{Goals}\label{subsec:Goals}

In this project, we perform a large-scale survey of L- and T-type brown dwarf atmospheres by fitting $\sim$300 archival spectra with theoretical models. This project differs from most other model-fitting surveys in extent. We know of only one other survey where model fits were performed for such a large sample of L- and T- dwarf spectra. 

In their paper, \citet{sanghi2023hawaii} use model fits to a sample of 1054 brown dwarfs of types M6-T9 to derive relationships between age, bolometric luminosity, temperature, surface gravity, and spectral type. However, they use the ATMO 2020 \citep{2020A&A...637A..38P} and BT-SETTL \citep{Allard_2012, 2015A&A...577A..42B} model sets. For our analysis, we instead use models from the Sonora and Phoenix groups. Comparing and contrasting our results with those of \citet{sanghi2023hawaii} can help to illuminate the relative strengths and weaknesses of the four different model sets.

Our goals are to evaluate how well the models are fitting real spectra and to better quantify how the properties of brown dwarf atmospheres such as temperature and cloudiness change with spectral type. We also seek to identify "deviant" objects: brown dwarfs with spectral features that are notably different from other objects of the same spectral type. 

{In order to validate our results, we compare our atmosphere model fits to the parameters derived from evolutionary models in \cite{sanghi2023hawaii}. }

\section{Data}\label{sec:Data}

\begin{figure}[htb!]
    \plotone{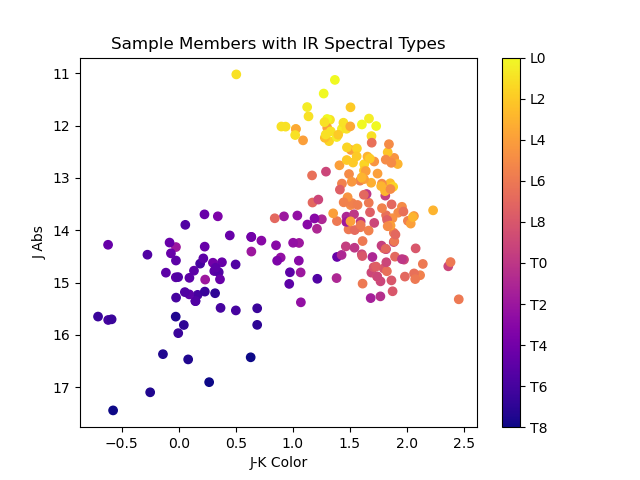}
    \caption{Absolute J$_{2MASS}$ magnitude vs (\textit{J-K})$_{2MASS}$ color of all the sources in the sample with published parallaxes and infrared spectral types. The data points are color-coded by infrared spectral type.}
    \label{fig:JvsJK}
\end{figure}

\subsection{Archival Spectra}\label{sec:archivalSpectra}

The archival spectra were downloaded from the SpeX Prism Library (SPL) \citep{burgasser2014spex}. When the library contained multiple spectra for the same source, only the most recent spectrum with the highest resolution was used. This left 301 spectra of unique sources with spectral types between L0 and T8. These archival spectra were obtained by multiple groups and published in various papers from 2004-2010. The full list of objects and their sources is given in Table \ref{tab:speXData}. 

The SPL spectra were all taken with the SpeX spectrograph \citep{Rayner_2003} at the 3.0 m NASA Infrared Telescope Facility (IRTF). The spectra are low resolution ($\lambda / \Delta \lambda \sim 75-200$) and span the near infrared from about 0.8-2.5 $\mu m$.

\textit{H} and \textit{Ks} magnitudes from the Two Micron All-Sky Survey (2MASS) were used to flux-calibrate each spectrum \citep{2006AJ....131.1163S}. We chose not to use the \textit{J$_{2MASS}$} magnitude due to the large potential for error that Earth's atmosphere introduces to the water band around 1.15 $\mu m$ for observations from the ground. \textit{H$_{2MASS}$} and \textit{Ks$_{2MASS}$} scale factors were found for each archival spectrum by convolving the spectrum with the corresponding filter profile and calculating the flux ratio between the resulting spectrum and known magnitude. The final scale factor used for each object was the average of its \textit{H$_{2MASS}$} and \textit{Ks$_{2MASS}$} scale factors.

\subsection{Parallaxes}
We found published parallaxes for $\sim$170 of our 301 sample objects on the open-source SIMPLE Database. An additional $\sim$70 objects had published parallaxes listed on SIMBAD \citep{Wenger_2000}. Some objects had multiple published parallaxes. In this case, the parallax with the smallest uncertainty was used. The majority of these parallaxes came from Gaia Early Data Release 3 \citep{2016,2021} or \cite{Best_2020}. The individual parallaxes and references for each object are given in Table \ref{tab:speXData}. 

We used the parallaxes to convert the apparent \textit{J$_{2MASS}$} magnitudes to absolute \textit{J$_{2MASS}$} magnitudes. Figure \ref{fig:JvsJK} shows the absolute \textit{J$_{2MASS}$} magnitudes, \textit{J-K} colors, and infrared spectral types for all objects in the sample with published parallaxes.

\subsection{Spectral Types}
The spectral types used in this project are those listed in the SIMPLE database. We chose to differentiate between optical \citep{Kirkpatrick_1999} and infrared \citep{burgasser2001classification, Geballe_2002} spectral types because these are assigned based on different spectral indices and thus do not always agree. In fact, a large difference in optical and infrared classification can be insightful; a significant discrepancy between the two types for the same object could possibly indicate binarity, variability, inhomogeneous cloud-clearing, or some other interesting physical process occurring in the atmosphere of the object. The spectral types for each object and their references are given in Table \ref{tab:speXData}.

\section{Models}

\begin{deluxetable}{cccccc}[htb!]
\label{tab:models}
\tablecaption{Model Parameters}
\tablehead{\colhead{Model Set} & \colhead{Temperature} & \colhead{log($g$)} & \colhead{log $K_{zz}$} & \colhead{Clouds} & \colhead{Metallicity}} 
\startdata
Bobcat & 200-2400K & 3-5.5 & None & None & -0.5,0.0,+0.5 \\
Cholla & 500-1300K & 3.5-5.5 & 2,4,7 & None & Solar \\
Diamondback & 900-2400K & 3.5-5.5 & None & $f_{sed}$ = 1,2,3,4,8,nc & -0.5,0.0,+0.5 \\
Phoenix & 800-2000K & 3.5-5.5 & 4,8 & $a_0$ ($\mu$m) \& $P_c$ (bar) & Solar \\
\enddata
\tablecomments{A summary of the parameter options for the model sets used in this project. More details can be found in the presentation papers for each model set \citep{Marley_2021,Karalidi_2021,morley2024sonora,Mukherjee_2023,Brock_2021}.}
\end{deluxetable}

We chose to use forward model sets from both the Sonora and Phoenix families so that we could investigate their relative strengths and weaknesses in fitting the archival spectra through the L/T sequence. In the next two sections, we give a brief overview of the characteristics of these model sets. We refer readers to the presentation papers of each model set for a more detailed description. Table \ref{tab:models} gives a summary of the main parameter grids of the different model subsets.

\subsection{Sonora Models}
The general parameters used in the Sonora model grids are temperature, surface gravity, eddy diffusion coefficient, sedimentation efficiency, metallicity, and C/O ratio. 

There is not currently a single model suite from the Sonora group that accounts for both clouds and disequilibrium chemistry while also covering a broad enough temperature range to fit the entire L/T sequence. Instead, there are three different Sonora model subsets--the Bobcat, Cholla, and Diamondback models-- that each have some of the needed properties. We combined all three of these subsets into one master suite of Sonora models and used this set for our model fits.

The Sonora Bobcat models \citep{Marley_2021,ZenodoBobcat} assume cloudless atmospheres in rainout chemical equilibrium. The temperature grid runs from 200-2400 K. The grid spacing is 25K from 200-600 K, 50 K from 600-1000 K, and 100 K from 1000-2400 K. The surface gravity grid covers log($g$) 3-5.5 with a step size of 0.25. Metallicity ([M/H]) can take values of -0.5, 0.0, or +0.5. The models with [M/H]=0.0 and log($g$) =5 also have C/O options of 0.5, 1.0, and 1.5, while all other models only have C/O = 1.0.

The Sonora Cholla models \citep{Karalidi_2021,Mukherjee_2023,ZenodoCholla} also assume cloudless atmospheres, but allow for disequilibrium chemistry. The Cholla temperature grid covers 500-1300 K with a constant spacing of 50 K. The surface gravity grid has a step size of 0.25 and ranges from log($g$) 3.5-5.5. All models have solar [M/H]. The extent of vertical mixing is quantified via the eddy diffusion parameter (log $K_{zz}$) with values of 2, 4, or 7 (cgs).

The Sonora Diamondback models \citep{morley2024sonora} assume cloudy atmospheres in chemical equilibrium. The temperature grid runs from 900-2400 K with a constant step size of 100 K. Surface gravity values are log($g$) 3.5-5.5 with a step size of 0.5. [M/H] can take values of -0.5, 0.0, or +0.5, and C/O is always solar. Cloud properties are adjusted with a sedimentation efficiency parameter $f_{sed}$ \citep{Ackerman_2001,Saumon_2008} that has options of 1, 2, 3, 4, 8, or nc ("no clouds"). The $f_{sed}$ value corresponds to how efficiently individual cloud grains are able to grow via sedimentation. Small $f_{sed}$ values mean small individual cloud grains, while large $f_{sed}$ values mean that the grains are able to grow larger and heavier. However, larger and heavier grains more quickly fall out of the upper atmosphere than smaller grains,{ thinning out the clouds both in terms of opacity and vertical extent.} Thus, small values of $f_{sed}$ correspond to thick, hazy cloud decks {that extend higher into the atmosphere and are }composed of small individual grains, while larger values of $f_{sed}$ correspond to thin, wispy clouds that {sit deeper in the atmosphere and} are composed of larger individual grains.

\subsection{Phoenix Models}
We chose to use the suite of models presented in \cite{Brock_2021} because this set was specifically targeted at replicating the cloud properties of brown dwarfs through the L/T transition. The temperature grid runs from 800-2000 K with a step size of 100 K. Log($g$) options range from 3 to 5.5 with increments of roughly 0.25, although values \textless3.5 are only available for temperatures \textless1200 and the quarter-increment options are not always provided for every temperature. Log $K_{zz}$ options are 4 and 8. All models have solar metallicity. 

This model suite implements the cloud model discussed in \cite{Barman_2011}. Instead of the $f_{sed}$ value used in the Sonora models, these models control cloud properties by varying the mean grain size $a_o$ and a pressure parameter $P_c$. The model assumes that the grain size distribution is log-normal and does not depend on atmospheric depth. The mean size $a_o$ can take values 0.1 to 10 $\mu$m. $P_c$ gives the pressure at which the condensate number density of the cloud begins to fall off (i.e., the vertical extent of the cloud), and can take values ranging from 0.1 to 2000 bar. {Note that not every $P_c$/$K_{zz}$ combination is available for every temperature/log($g$) combination.}

\subsection{Preparing the Models}
Because the resolution of the models is greater than that of the SpeX data, all of the models were smoothed using a boxcar smooth algorithm. A cubic spline function was then used to regrid the models to match the SpeX wavelength grid.

\section{Spectral Fitting}
We wrote a Python script to find the best-fit theoretical model for each archival spectrum. The code used the Nelder-Mead optimization function from the SciPy library \citep{2020SciPy-NMeth} to optimize the goodness-of-fit statistic $G_K$ \citep{Cushing_2008} for each model. The model with the lowest minimum $G_K$ was selected as the best-fit model.

Equation \ref{eq:Gk} gives the definition of $G_K$. The variables $f_i$, $F_{k,i}$, and $\sigma_i$ refer to the flux density of the data, flux density of the $k$th model, and error in the flux densities, all given at the $i$th wavelength. 

The SpeX archival data files occasionally had a few data points with recorded errors of $\sigma$ = 0, which would have resulted in division by zero if used in the calculation of $G_K$. For cases where a data file had fewer than 10 points with $\sigma$ = 0, we replaced these values with the average of the errors of the two points on either side. Because there were so few affected points, we do not believe that this adjustment impacted our fit results. There were six objects with significantly more points with $\sigma$ = 0. For these objects, we threw out the error term entirely. When possible, we flag these objects in our plots to make clear that the uncertainties in their fits are unknown.

The variable $w_i$ is a weight term that can be used to adjust for variable resolution across the spectrum. Because the SpeX data covers a short wavelength range with a fairly constant grid spacing, we set $w_i$=1 for all points. The constant $C_k$ is an unknown scaling value that is found by the optimization algorithm for each model. $C_k$ is equivalent to $(R/d)^2$. 

\begin{equation} 
    G_k = \sum_{i=1}^n w_i \left( \frac{f_i-C_kF_{k,i}}{\sigma_i} \right)^2
    \label{eq:Gk}
\end{equation}

We ran the fitting algorithm twice for each archival spectrum: once using the Sonora model sets and again using the Phoenix models. We recorded the parameters of the best-fit models for each archival spectrum and used these parameters as estimates for the physical properties of the target brown dwarf. Our results are presented in the next two sections.

In order to find uncertainty estimates for the best-fit parameters, we resampled each spectrum 100 times using the random.normal() function from the NumPy library--with the archival errors used as the input standard deviation--and fit each of the resulting spectra. We then found the mean and standard deviation of the best-fit parameters from all 100 fits. For many of the objects in our sample, the resampling was not enough to change the best-fit model for any of the 100 fits. For some of the fainter objects with lower signal-to-noise, this method did yield nonzero uncertainties, which are plotted as error bars in the figures below.

This method likely underestimates the uncertainty due to the coarseness of the forward model grids. However, we chose not to interpolate between models because recent studies have shown that interpolation can be dangerous for models of cool objects \citep{wheeler2024korgfittingmodelatmosphere}. Instead, we just note that, unless otherwise specified, the uncertainties of the parameter fits are smaller than the resolution of the forward model grid.

\begin{figure}[htb!]
    \plottwo{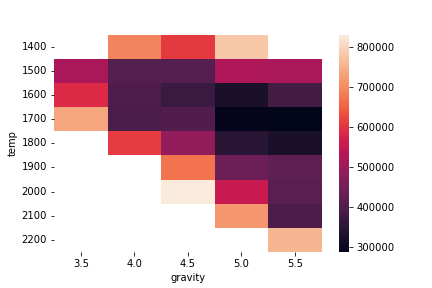}{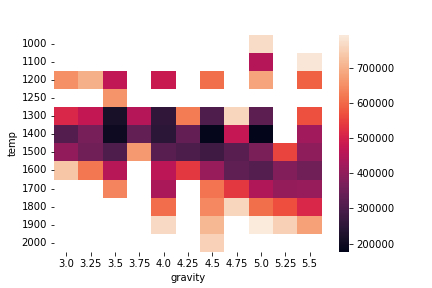}
    \caption{Two sample heat plots showing the distribution of the goodness of fit statistic $G_K$ in temperature-gravity space. Darker colors correspond to lower values of $G_K$ and thus better model fits. The heat plot on the left corresponds to the top Sonora model fits to 2MASSW J0036159+182110 (spectral type L4). The fit was reasonably well constrained in terms of model temperature and gravity, with the best model fit giving a temperature of 1700 K and log($g$) of 5.5. The heat plot on the right shows the top Sonora model fits to IPMS J013656.57+093347.3 (spectral type T2.5). This object is a member of the Overluminous T family described in Section \ref{subsec:overluminousT} because it is brighter in the \textit{J}-band than is typical for an object of type T2. The heat map shows that log($g$) is poorly constrained by model fits to this deviant object.}
    \label{fig:heatmap}
\end{figure}

We also created heat maps showing the distribution of $G_K$ versus pairs of model parameters for each archival object as another method of expressing the confidence in the fits. The heat maps show at a quick glance how well different best-fit model parameters were constrained. Example heat maps, one map from a well-fit object and another map from a poorly-fit, "deviant" object (discussed in Section \ref{subsec:overluminousT}), are given in Figure \ref{fig:heatmap}. Due to space constraints, we will not present the heat maps for all 305 objects in our sample in this paper, but we have made them available on Zenodo: \dataset[doi:10.5281/zenodo.15314197]{https://doi.org/10.5281/zenodo.15314197}.

In order to investigate possible binarity, we generated composite models made from every possible combination of two theoretical models by summing together their two different flux values at each wavelength point. These composite models mimic the spectral signatures of unresolved spectral binary systems \citep{Burg10,Bard2014}. We then fit the archival spectra to these composite models using the same method described in the previous paragraphs. We compared the single and composite fits by-eye to determine whether the composite fits were significantly better able to replicate the archival spectra. Due to the time required, we only ran composite fits on objects that were poorly fit by single models. We only used the Sonora models for these binary fits because the Sonora models better constrained temperature 
 than the Phoenix models (discussed in Section \ref{subsec:relationships}). We also removed the Bobcat models from the Sonora suite when running the binary fits because they so rarely beat out the Diamondbacks that their inclusion was not worth the exponential increase in code run time.

\section{General Trends}

\subsection{Distribution of Sonora Fits}
\begin{figure}[htb!]
\gridline{\fig{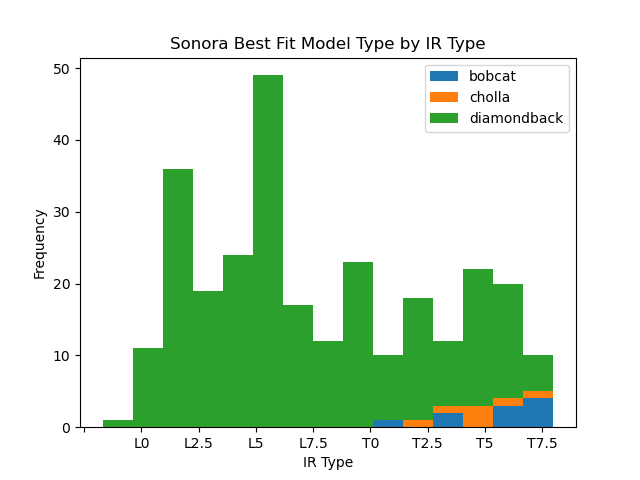}{0.3\textwidth}{(a)}
          \fig{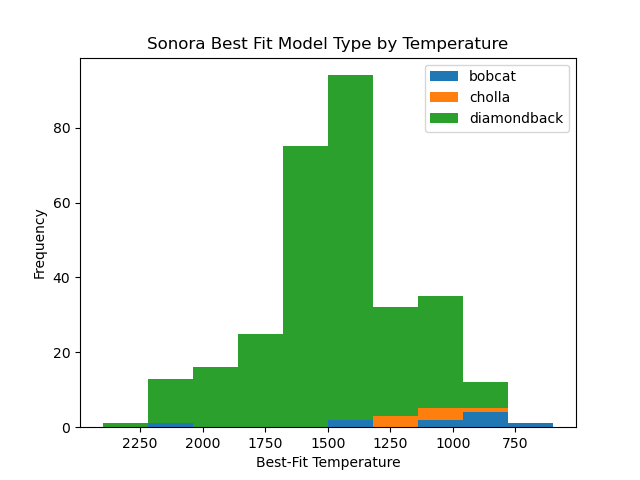}{0.3\textwidth}{(b)}
          \fig{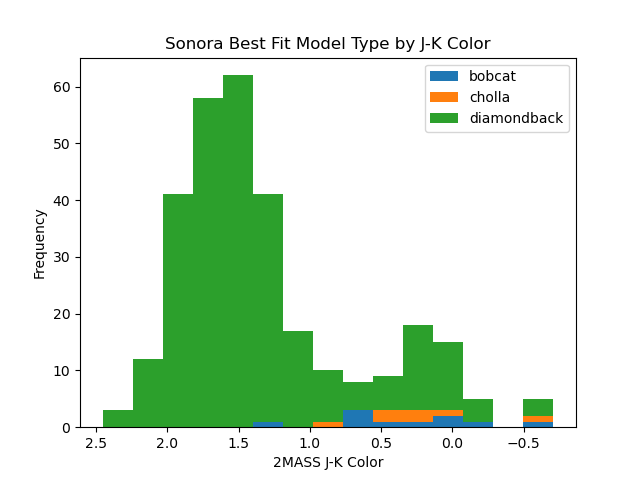}{0.3\textwidth}{(c)}}
\caption{Histograms showing the distribution of objects that were best-fit by Bobcat (blue), Cholla (orange), or Diamondback (green) models. The plots show the distribution by (a) infrared spectral type, (b) best-fit temperature, and (c) $(J-K)_{2MASS}$ color index. Nearly every object in our sample was best-fit by a Diamondback model.}
\label{fig:sonoraModelDistribution}
\end{figure}

Because of the different cloud and temperature properties of the three Sonora model subsets, we expected there to be spectral type regimes that were predominantly better-fit by either Bobcat, Cholla, or Diamondback models. Plots of the distribution of best-fit model type versus infrared spectral type, best-fit temperature, and \textit{J-K} color are given in Figure \ref{fig:sonoraModelDistribution}.

The objects in our sample were all best fit with Diamondback models, almost without exception. Because the Bobcats are the oldest of the Sonora subsets and include neither disequilibrium chemistry nor clouds, it was not a surprise that Bobcat models were almost never the best fits. However, we expected the Cholla models to take over after the L/T transition when the 8-10 $\mu$m silicate cloud feature disappears \citep{suarez22} and the effects of disequilibrium carbon chemistry become noticeable. Instead, the cloudy, equilibrium Diamondback models continue to dominate over the cloudless, disequilibrium Cholla models all the way up until 900 K when the Diamondback temperature grid bottoms out. This suggests that---at least in the near-infrared---the effects of clouds are more important than the effects of disequilibrium chemistry. The fact that the Bobcats show up in the fits to cooler objects about as frequently as the Chollas again suggests that disequilibrium chemistry is not very impactful in the near-IR.

\subsection{Temperature}\label{subsec:relationships}
We next plotted the best-fit model parameters against each other in order to investigate how they might be related. We note here that for all plots involving spectral type, we looked at the infrared and optical types independently to account for differences between the two methods of classification. In most cases, the plots and resulting analysis were nearly identical for the two spectral typing systems. To conserve space, we will only include the infrared spectral type plots in this paper. 

\begin{figure}[htb!]
    \plottwo{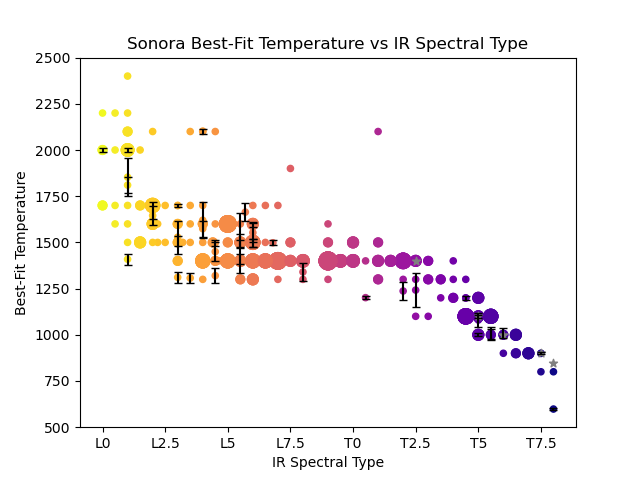}{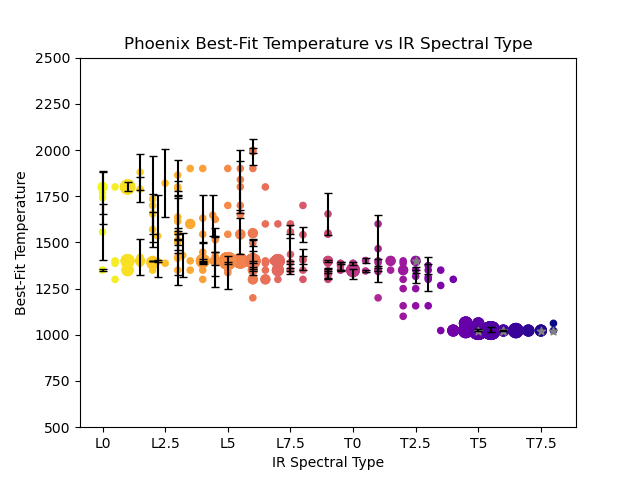}
    \caption{Plots of best-fit Sonora (left) and Phoenix (right) model temperature vs infrared spectral type. The markers are color-coded by spectral type. The size of the markers corresponds to the number of objects represented by each point. The temperatures plotted are the mean temperatures yielded after 100 MC resamples. Error bars come from the standard deviation of all 100 fits and are only shown when nonzero. The six grey star markers denote objects in the sample with unknown errors and thus undetermined uncertainty to their fit temperatures. The Sonora temperatures generally decrease, but plateau through the L/T transition. The Phoenix temperatures suggest a similar trend, but the scatter in temperature through the L types as well as the limited temperature range make the plateau less clear.}
    \label{fig:tempSpecType}
\end{figure}

\subsubsection{Temperature vs Spectral Type}

Plots of best-fit temperature according to the Sonora and Phoenix model fits versus infrared spectral type are given in Figure \ref{fig:tempSpecType}. For the Sonora temperatures, the general shape of both the optical and infrared spectral-type plots mimic a cubic function: temperatures quickly decrease with spectral type until reaching the L/T transition, where the temperature plateaus. The temperature then falls off again after the L/T transition. This plateau at the L/T transition suggests that clouds are clearing away or sinking below the photosphere, allowing light from deeper, hotter levels of the atmosphere to escape and causing the effective temperature to appear to hold steady for a time \citep{Burgasser_2002, Saumon_2008, Kirkpatrick_2021}. 

There are five notable outliers with Sonora best-fit temperatures that are much hotter than other objects of the same infrared spectral type. The most extreme outlier is SDSS J015141.69+124429.6, with an IR spectral type of T1 \citep{Burg06} but a best-fit temperature of 2100 K. The model fit to this object is poor, likely due to its abnormally flat \textit{H}-band peak. This object is discussed in greater detail in Section \ref{sec:interestingfamilies}. 

SDSS J112118.57+433246.5 also sticks out with its IR spectral type of L7.5 pec \citep{Chiu06} and best-fit temperature of 1900 K. In a later paper, \cite{schn2014} assign an earlier spectral type of L5 to this object. Even as an L5, this object sticks out as being too hot for its spectral type. The model fit is good through the H and acceptable through the \textit{K} band, but the \textit{J} band is markedly overfit. We ran a binary fit to this object, and found that a composite model with component temperatures of 1800 K and 1600 K can help to rectify much of the \textit{J} band overfitting while tightening up the \textit{K}-band fit. \cite{gag15} also flag this object as a visual spectral binary candidate, although they estimate the primary/secondary spectral types to be L6.8/T5, later than our component temperatures suggest.

The last three notable outliers--2MASS J09211410-2104446 (optical L1.5/IR L4), SDSS J104842.84+011158.5 (optical L1/IR L4), and 2MASS J14313097+1436539 (optical L2/IR L3.5)--are clustered together with best-fit temperatures of 2100 K. All three objects show a difference between their optical and IR classifications of at least 1.5 types. Their optical types are early enough to explain a hot temperature of 2100 K, but their IR types place them at least 400K warmer than the next hottest objects with the same classification. Both 0921-2104 and 1048+0111 have been observed to be variable \citep{Koen13} and to give off H$\alpha$ emission \citep{mclean12,rein08}, but 1431+1436 did not appear in the same studies. The three spectra have marked differences from each other, suggesting that their hot fits might not all have the same cause, even though they are clustered together in temperature-IR spectral type space. Our binary fits to 0921-2104 and 1431+1436 did not visually show enough improvement over the single fits to warrant flagging them as possible binaries. However, the binary fit to 1048+0111 (primary of 2100 K/ secondary of 1300 K) did show clear improvement to the \textit{K} and \textit{H} bands. 

The Phoenix temperature distribution also shows a loosely decreasing trend with spectral type. The MC resampling yielded higher uncertainties on average--particularly for L-type objects--than were seen in the Sonora temperature results (see the error bars in Figure \ref{fig:tempSpecType}). Like the Sonora fits, the Phoenix temperatures plateau around 1400 K through the L/T transition. However, the Phoenix temperatures show more scatter in the mid-to-late Ls than was seen for the Sonora temperatures. There were too many objects with temperatures above the 1400 K plateau to identify and discuss each one individually here. 

At first glance, the collection of points at 1000 K suggests that the temperature grid of the Phoenix model suite used here does not extend low enough to comfortably cover the entire L/T sequence. However, the model suite includes temperatures down to 800 K. For some reason, models with temperatures below 1000 K are never the best fit, even though the Sonora fits suggest that the late T-type objects in our sample cool below 800 K. We note that this Phoenix model suite was designed to fit L/T transition-type atmospheres, specifically spectral types L3-T5. Thus, it might be best to disregard the Phoenix temperature fits to objects with spectral types outside this range. Because nearly all of the deviant objects discussed in Section \ref{sec:interestingfamilies} are L/T transition objects, we still found it informative to include this model suite in our analysis even with the limited temperature range.

\subsubsection{Temperature vs \textit{J-K} Color}

\begin{figure}[htb!]
    \plottwo{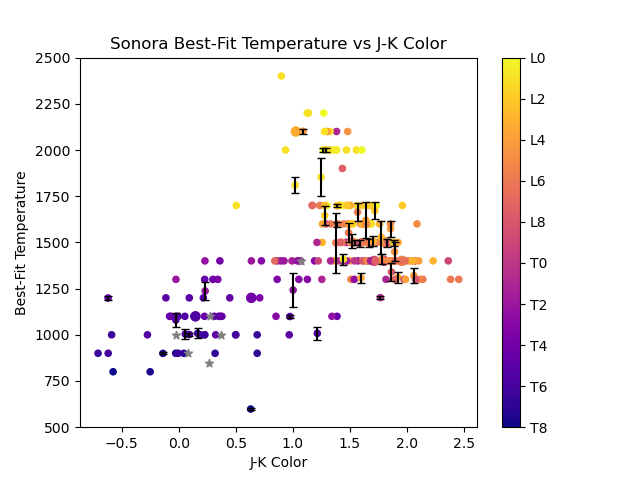}{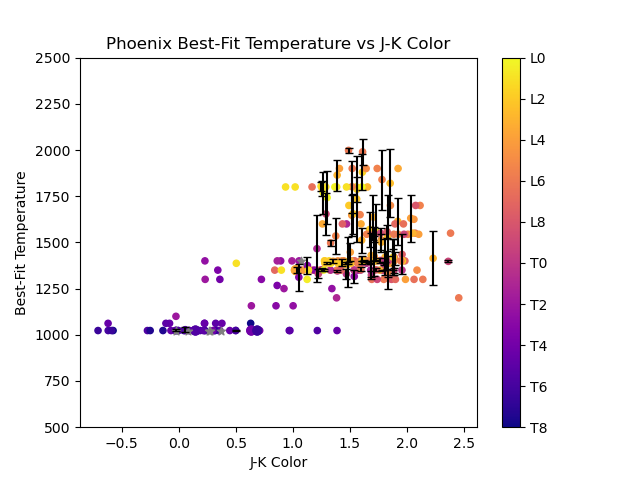}
    \caption{Plots of best-fit Sonora (left) and Phoenix (right) model temperature vs \textit{J-K} color, color-coded by infrared spectral type. Again, the temperatures and error bars come from the 100 MC resampling results, and only the nonzero errors are shown. The grey stars again denote the objects with unknown uncertainties. There is a trend of general reddening with later spectral type until the L/T transition, where colors quickly become bluer. The larger uncertainties and truncated temperature range make this trend more difficult to see for the Phoenix fits as compared to the Sonora fits.}
    \label{fig:tempColorType}
\end{figure}

We also compared the best-fit temperatures and \textit{J-K} colors for the objects in our sample. The Sonora model plot in Figure \ref{fig:tempColorType} shows a clear trend: the \textit{J-K} color reddens steadily until rapidly swinging towards bluer colors at the L/T transition, hinting once again that there is a major shift in cloud properties at the L/T transition. This \textit{J-K} color transition point aligns with the temperature plateau seen in Figure \ref{fig:tempSpecType}. 

The most notable outlier here is GJ 1048B with a best-fit temperature of 1700 K and a \textit{J-K} color of 0.504. This is significantly bluer than any other object of the same temperature or spectral type. 

The Phoenix plot in Figure \ref{fig:tempColorType} shows a similar trend, but with more scatter/greater uncertainty and the same temperature pileup at 1000 K seen in Figure \ref{fig:tempSpecType}. 

Theory suggests that L dwarfs have clouds composed of magnesium-silicate-oxides and iron metal in their upper atmospheres \citep{Lodders_2004}. These clouds thin, develop holes, or sink below the visible photosphere at the L/T transition and allow bluer light to escape from hotter depths of the atmosphere \citep{Burgasser_2002, Marley_2010}. At the same time, methane begins to condense in the upper atmosphere and resulting methane absorption removes light from the \textit{K}-band. Both of these effects together, the brightening of the \textit{J}-band and the diminishing of the \textit{K}-band, cause the \textit{J-K} color to swing to the blue at the L/T transition \citep{Burgasser_2006}.

\subsubsection{Comparison to SANGHI23}

\begin{figure}[htb!]
    \centering
    \plotone{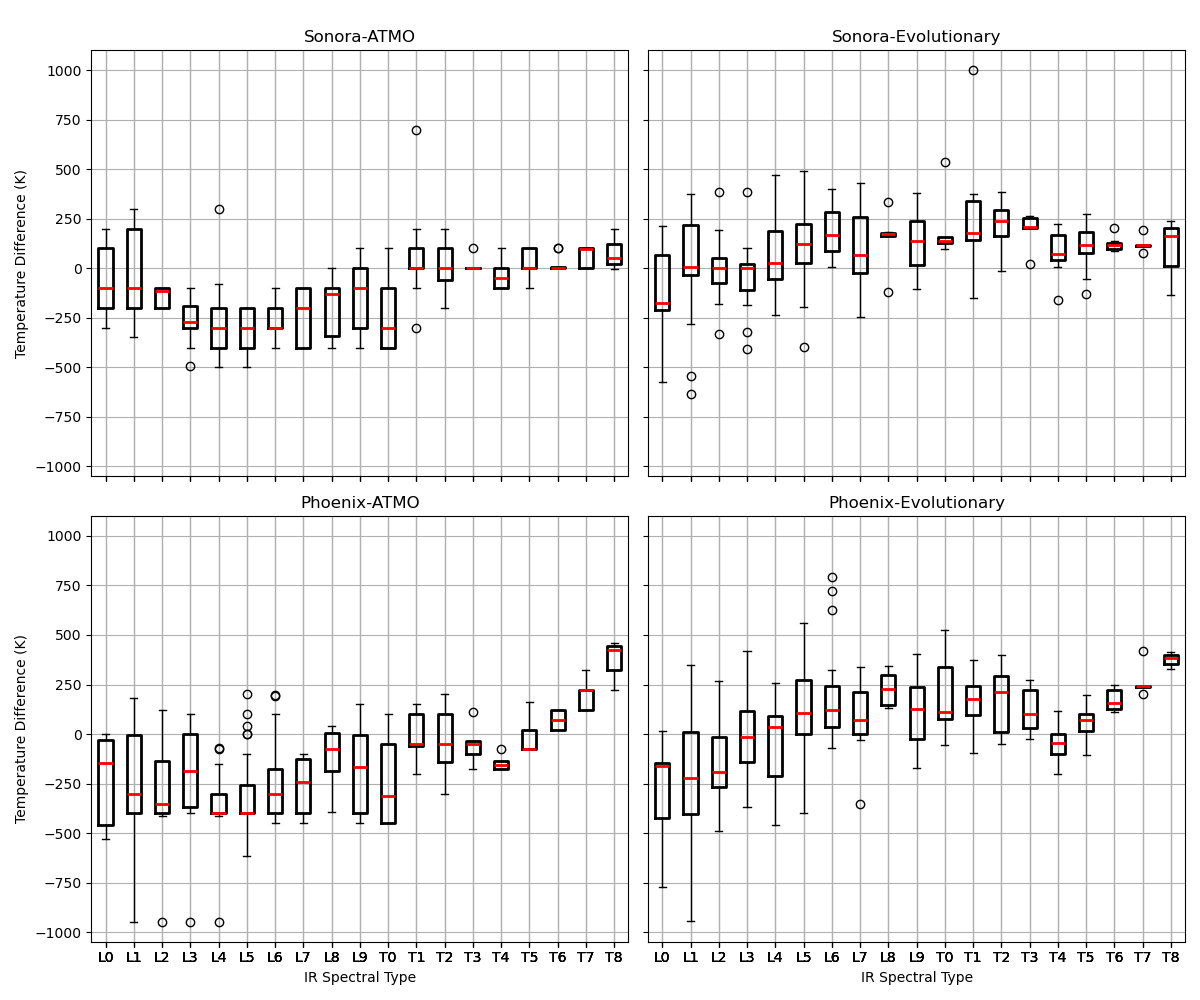}
    \caption{Box plots showing the distribution of the differences between the best-fit values of T$_{eff}$ found in this work (Sonora models on top and Phoenix models on the bottom) and those found in \cite{sanghi2023hawaii} for the 247 objects shared by both samples. In their paper, SANGHI23 find best-fit T$_{eff}$ using forward model fits to the ATMO models (left) and also derive T$_{eff}$ values using evolutionary models (right). In each case, the SANGHI23 temperatures are subtracted from the temperatures found in this work.}
    \label{fig:teffs}
\end{figure}

{In order to benchmark our best-fit values for T$_{eff}$, we compare them to the evolutionary-model-derived T$_{eff}$ values given in \citet{sanghi2023hawaii} (SANGHI23). We note that the authors used evolutionary models from both \citet{Saumon_2008} and \citet{2015A&A...577A..42B} in order to span the full range of luminosities included in their sample. The full description of how the authors selected between the model sets for each object is given in Section 7 of their paper.}

{The sample from their paper included 247/301 of the objects from our sample. In Figure \ref{fig:teffs}, we plot the residuals between the SANGHI23 evolutionary and our best-fit Sonora and Phoenix T$_{eff}$ values for the 247 common objects.}

{The Sonora T$_{eff}$ values align reasonably well with the evolutionary values. With the exception of L0, the mean difference for each spectral type bin is always between 0 and 250 K. The Sonora T$_{eff}$ values are generally higher than the evolutionary values.}

{Between L3-T5, the Phoenix T$_{eff}$ values also agree with the evolutionary values fairly well, with mean differences between -50 and 250 K. For these spectral types, the Phoenix temperatures are also typically higher than the evolutionary T$_{eff}$ values. Before L3 and after T5, the Phoenix T$_{eff}$ values diverge more noticeably from the evolutionary values. This is unsurprising, as we have already established that the Phoenix models used in this paper are best suited for types L3-T5.}

{SANGHI23 also perform traditional forward model fits to a combined model suite containing the cloudy BT-SETTL \citep{Allard_2012,2015A&A...577A..42B} and cloudless ATMO 2020 models \citep{phillips2020}. We follow the convention used in their paper and refer to their combined suite as simply the "ATMO" models. We display the differences between our best-fit T$_{eff}$ values and the SANGHI23 ATMO best-fit T$_{eff}$ values in the left panels of Figure \ref{fig:teffs}.}

{The Sonora and ATMO values agree very closely for spectral types $\geq$ T1. Through the L types, the Sonora temperatures are generally lower than the ATMO values, with the mean difference for each bin ranging from about -100 to -300 K. A closer look at the distribution of the ATMO fits shows that BT-SETTL models are always the best fit for spectral types L0-T1, types T3.5-T9 always match with ATMO 2020 models, and spectral types T2-T3.5 can be fit by either model set. The transition from negative mean difference to a mean difference of $\sim$0 K (T1) does not correspond to the spectral type where ATMO 2020 models begin to appear in the fits (T2), suggesting that the change in mean temperature difference has more to do with the physical characteristics of L vs T-type atmospheres than the differences between the ATMO 2020 and BT-SETTL model sets.}

{The Phoenix values are also generally lower than the ATMO values through the L types, with mean differences between about -50 and -375 K. The increasing mean difference in the late T-types again is likely due to the Phoenix models being designed for spectral types L3-T5.}

{Using their evolutionary models as the benchmark, SANGHI23 observe that the ATMO models underestimate the temperature of objects in M/L transition. They hypothesize that this discrepancy is due to insufficient dust inclusion in the BT-SETTL models. The Sonora and Phoenix models both give temperatures that are generally even lower than the ATMO values for M/L transition objects. Following the logic of SANGHI23, it is possible that these model sets also might need to increase the dust opacity for their warmer models.}

{SANGHI23 also find that the ATMO models tend to overestimate the temperatures of late L dwarfs. They attribute this to an overestimation of the impact of dust opacity for these cooler models. This could explain part of the difference between the temperatures in this work and the SANGHI23 ATMO temperatures for late L dwarfs. However, Figure \ref{fig:teffs} shows that the Sonora and Phoenix models both also overestimate late L temperatures relative to the evolutionary models, although to a lesser degree than the ATMO models. Thus, it is possible that the Sonora and Phoenix models also need slightly lower dust opacity for late L-type temperatures.}

\subsection{Surface Gravity}
\begin{figure}[htb!]
    \plottwo{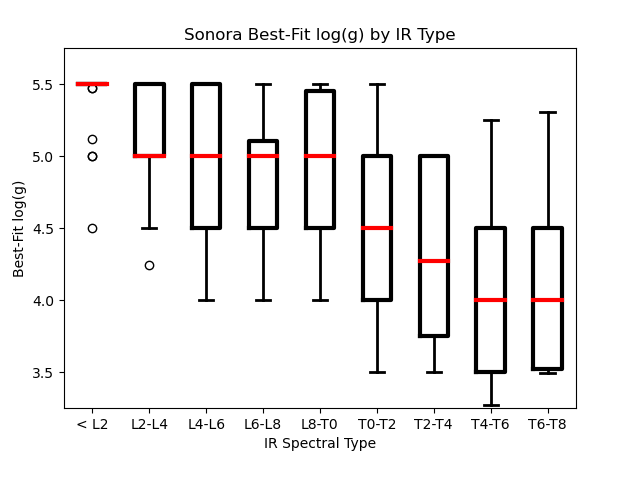}{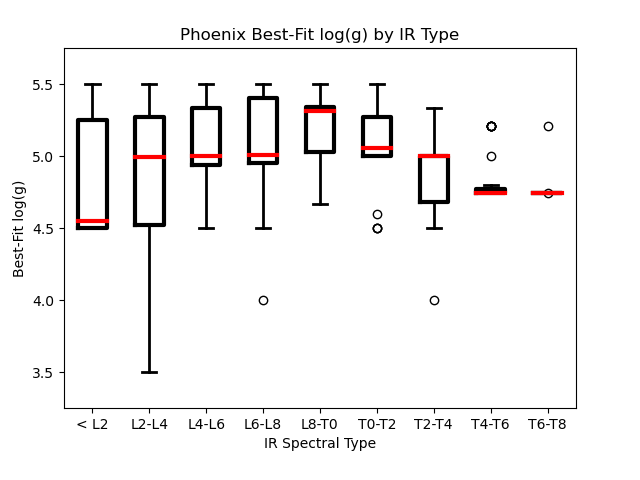}
    \caption{Box plots showing the distribution of best-fit gravity vs infrared spectral type for the Sonora (left) and Phoenix (right) model fits. The mean value for each spectral type bin is given in red. The two model sets give entirely different trends in log($g$), suggesting that surface gravity fit results to the near-infrared are model-dependent and poorly constrained.}
    \label{fig:gravity}
\end{figure}

Figure \ref{fig:gravity} gives the distribution of best-fit surface gravity versus infrared spectral type for both model sets. 

The mean log($g$) values from the Sonora model fits generally decrease through the L-T sequence. While there is a large scatter in best-fit log($g$) for each spectral type bin, log($g$) values \textless 4.0 do not appear until T0. The highest gravity value, log($g$) = 5.5, vanishes at the same spectral type. 

Although the Phoenix gravity grid spans the same range as the Diamondback grid, the Phoenix fits do not show the same decrease in mean log($g$) with spectral type. Instead, the mean log($g$) increases slightly with spectral type until T0 and then decreases again. The Phoenix models rarely yield best-fit log($g$) \textless 4.5. The only similarity between the Sonora and Phoenix results is that log($g$) = 5.5 once again vanishes at T0.

Young brown dwarfs have low-density, fluffy atmospheres that gradually shrink and condense after formation. Because of this, surface gravity can be used as a loose indication of age, with higher surface gravity values generally corresponding to older brown dwarfs \citep{Cruz_2009, Allers_2007}. 

Because the brown dwarf spectral types give an evolutionary sequence, we might expect to see a higher percentage of low-gravity/young objects in the L spectral types as compared to the T types. However, the Sonora results show the opposite trend while the Phoenix results rarely yield any low gravity values at all.

If the Sonora surface gravity best-fit values are accurate, there are very few young objects in our sample, and most of these objects are T-types. On the other hand, the Phoenix surface gravity values suggest that the only very young objects are early L-types. However, multiple objects of both L and T types in the sample have been previously flagged as young objects in the literature (see the spectral types in Table \ref{fig:gravity}). These discrepancies suggest that gravity is not particularly well-constrained by fits to the entire near-infrared spectrum. The results of \cite{Stephens_2009} also showed a difficulty to constrain log($g$) unless fits were restricted to particularly gravity-sensitive regimes.

\subsubsection{Comparison to SANGHI23}

\begin{figure}[htb!]
    \centering
    \plotone{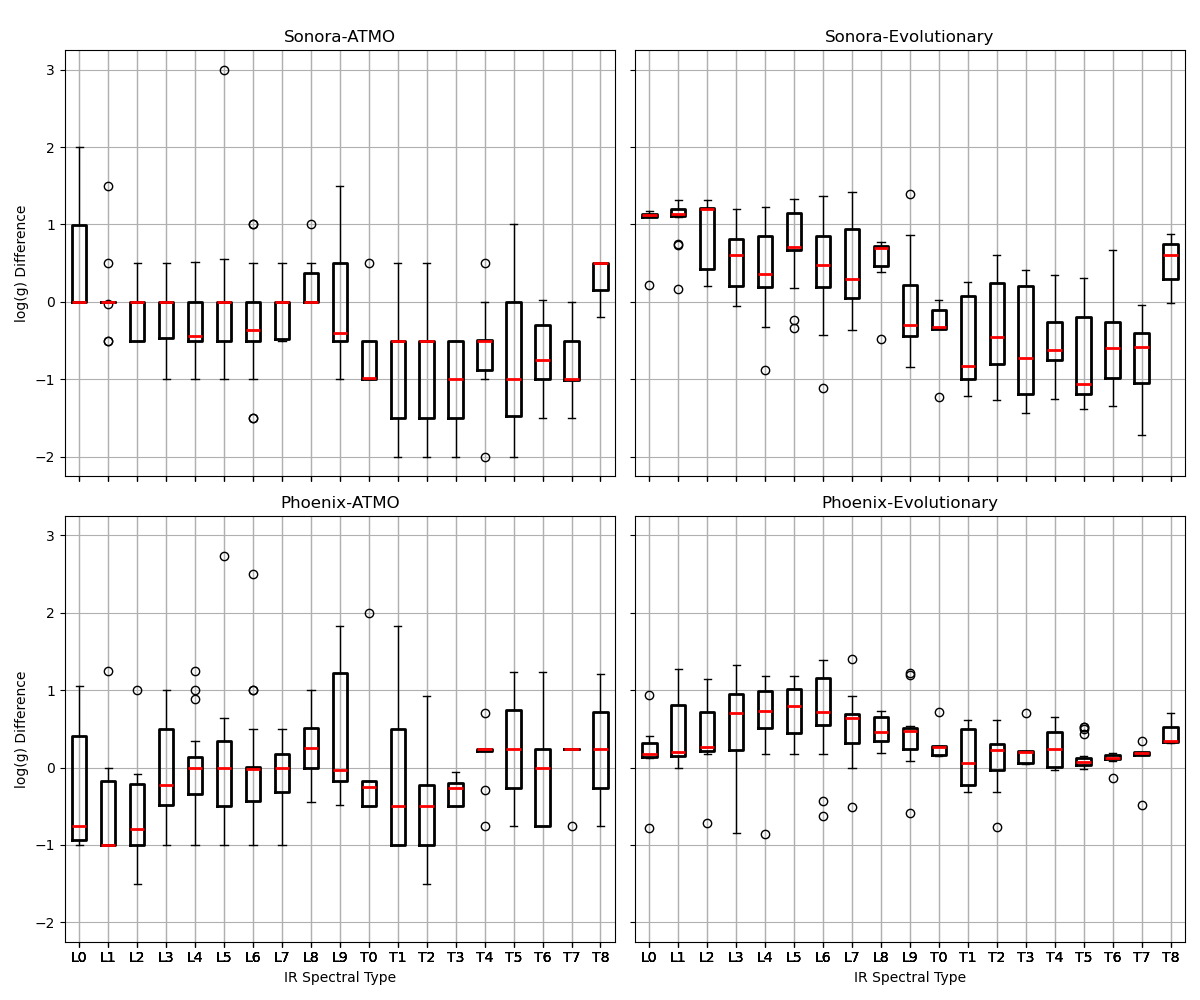}
    \caption{Box plots showing the distribution of the differences between the best-fit values of log(\textit{g}) found in this work (Sonora models on top and Phoenix models on the bottom) and those found in \cite{sanghi2023hawaii} for the 247 objects shared by both samples. In their paper, SANGHI23 find best-fit log(\textit{g}) using forward model fits to the ATMO models (left) and also derive log(\textit{g}) values using evolutionary models (right). In each case, the SANGHI23 log(\textit{g}) values are subtracted from the values found in this work.}
    \label{fig:loggs}
\end{figure}

{In Figure \ref{fig:loggs}, we compare the best-fit log($g$) values from this work with those derived in SANGHI23. }

{The Sonora best-fit log($g$) values align poorly with the evolutionary values. Until L8, the Sonora values are higher on average than the evolutionary values. Starting with L8, the Sonora values are too low compared to the evolutionary values, with the exception of the T8 bin. }

{The Phoenix best-fit log($g$) values align better with the evolutionary values, although there is still plenty of scatter. The Phoenix values on average are 0.5-1 dex higher than the evolutionary models for types L3-L9, but the early Ls and all T-type bins have mean differences that fall within $\sim$0.3 dex. }

{The atmospheric model log($g$) values from SANGHI23 also differ significantly from the values in this work. The Sonora values for log(\textit{g}) on average align fairly well with the SANGHI23 ATMO values through the L types, but starting with T0 the mean difference ranges between -0.5 and -1 dex. Again, the spectral type where the mean difference notably increases (T0) does not line up with the spectral type where the best-fit model type switches from BT-SETTL to ATMO 2020 (T2), so we cannot attribute the shift purely to differences between the two different ATMO model sets. The Phoenix and ATMO log($g$) values also differ significantly, except for the mid L-types where the mean difference is $\sim$0 dex.}

{These results again suggest that the best-fit log($g$) value is poorly constrained/ highly model-dependent when using data that only spans the near-IR.}

\subsection{Metallicity}

\begin{figure}
    \centering  \plottwo{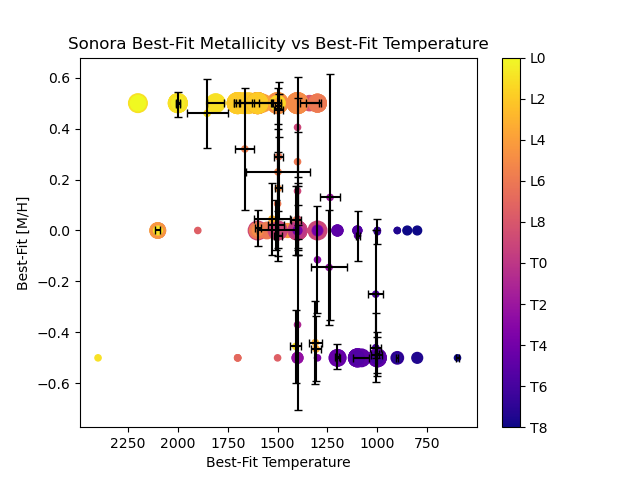}{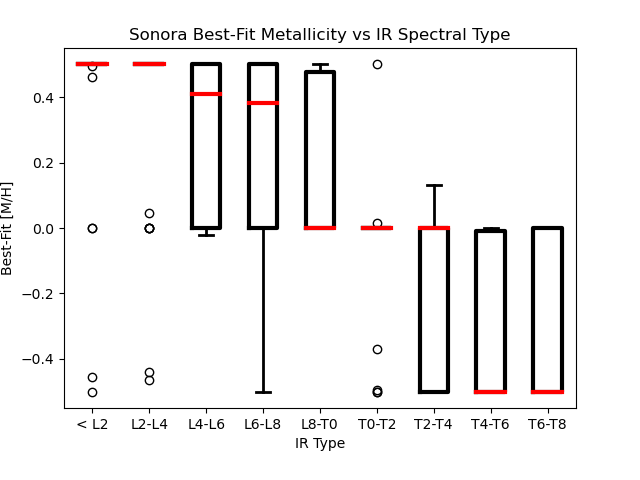}
    \caption{The left scatter plot shows the best-fit metallicity [M/H] vs best-fit temperature of the objects in our sample. The markers are color-coded by IR spectral type and sized based on the number of objects they represent. The values of [M/H] and temperature are averages after 100 MC resamples for each object. Uncertainties are also calculated based on the resampling results. The right bar plot shows the best-fit [M/H] vs IR spectral type. The mean [M/H] for each bin is depicted in red. Both plots show a decrease in [M/H] with decreasing temperature/later type.}
    \label{fig:metallicity}
\end{figure}

The Bobcat and Diamondback sets give three options for metallicity (-0.5, 0.0, and +0.5), while the Cholla and Phoenix sets only allow for solar metallicity (0.0). Since the objects in the sample were almost never best-fit by Cholla models, we can safely investigate the possibility of a trend in metallicity for the Sonora fits. 

Figure \ref{fig:metallicity} shows the best-fit metallicity vs spectral type and best-fit temperature for our sample. There is an apparent trend of decreasing metallicity with later spectral type/cooler temperature. This trend is reminiscent of the similarly decreasing Sonora best-fit log($g$) values shown in Figure \ref{fig:gravity}. While this decreasing trend in metallicity with spectral type makes little physical sense, it could convey important information about the models. 

One observed consequence of low-gravity atmospheres is a reduction in collisionally-induced absorption (CIA) of H$_2$ relative to higher-gravity atmospheres. Reduced CIA would impact the spectrum of an object such that its \textit{H}-band peak would be sharper and less light would be absorbed from its \textit{K}-band. \citep{Kirkpatrick_2006}

Objects with greater [M/H] would also likely experience less CIA compared to lower-metallicity objects, due to the lower ratio of hydrogen to other atomic species. Thus, the similarly decreasing trends in metallicity and log($g$) would actually have opposite impacts on the strength of CIA: decreasing log($g$) would serve to weaken CIA, while decreasing [M/H] would strengthen CIA. \citep{Saumon_2012} 

This degeneracy in the impacts of log($g$) and [M/H] on CIA could at least partially explain why the Sonora fits tend to choose high-gravity, high-metallicity models for early L dwarfs and low-gravity, low-metallicity models for late T dwarfs.

{We note that the BT-SETTL and ATMO 2020 models used in SANGHI23 are all solar metallicity, do not have parameterized cloud properties, and do not incorporate disequilibrium chemistry, and thus cannot be used to benchmark our results for the remainder of this section.}

\subsection{Clouds}\label{subsec:clouds}

\begin{figure}[htb!]
    \plottwo{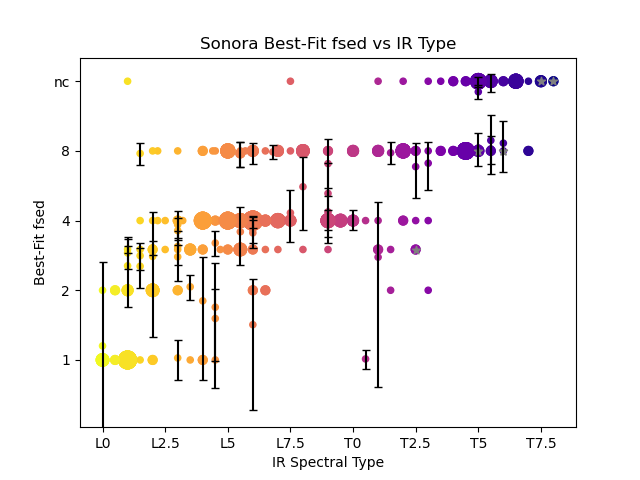}{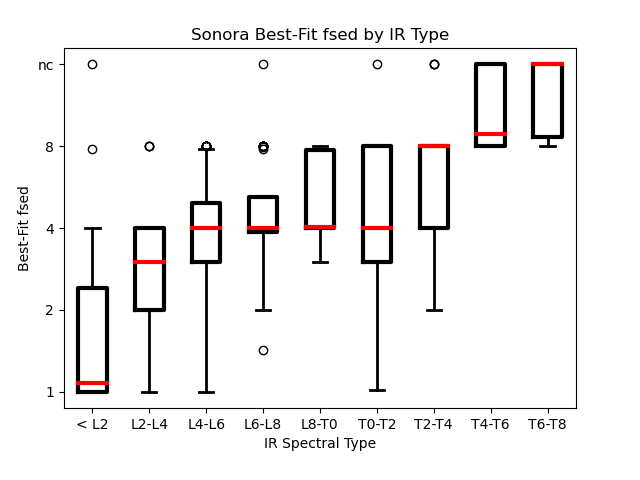}
    \plottwo{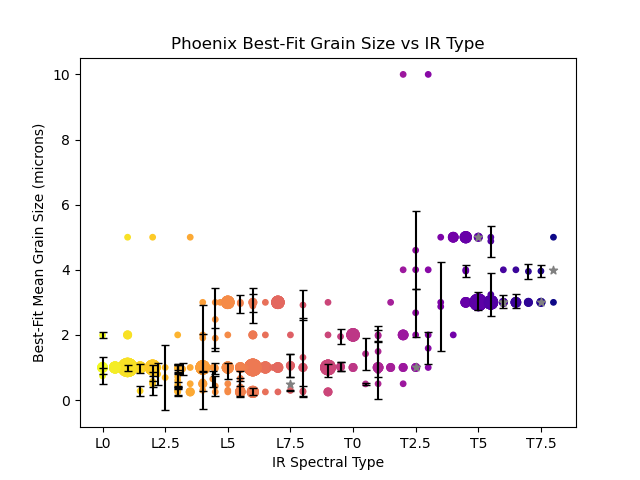}{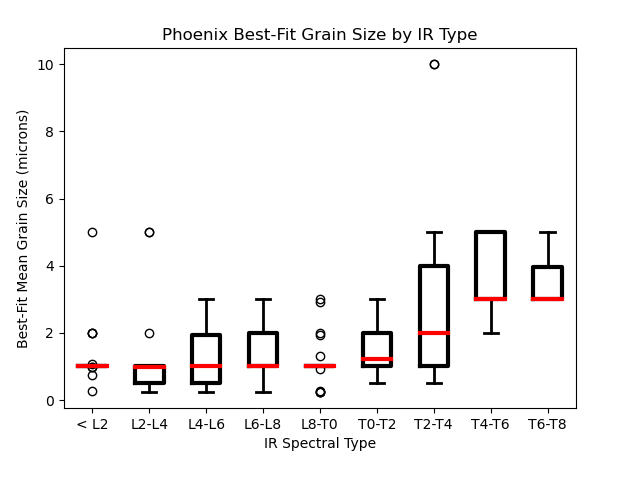}
    \plottwo{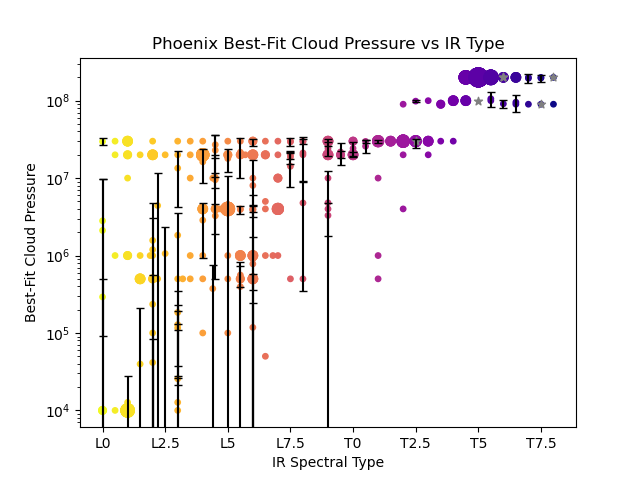}{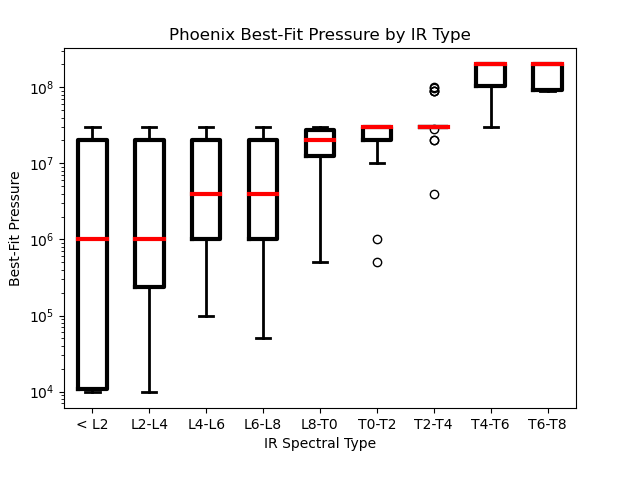}
    
    \caption{Scatter plots (left) and box plots (right) showing the distribution of the best-fit cloud parameters (fsed for the Sonora models and grain size/pressure at the top of the cloud deck for the Phoenix models) vs infrared spectral type. The scatter plot markers are color-coded by spectral type and sized according to the number of objects they represent. Parameter values and error bars were obtained by performing 100 MC resamples of each spectrum and finding the mean/standard deviation of the resulting fits. The increasing trend in all three parameters indicates {that silicate clouds fall deeper in the atmosphere through the L/T sequence, but are still high enough to influence the near infrared spectrum through the late T-types.}}
    \label{fig:clouds}
\end{figure}

The distribution of best-fit cloud parameters for the sample is given in Figure \ref{fig:clouds}. "Cloudiness" is represented by $f_{sed}$ for the Sonora models and by mean grain size ($a_0$) and the pressure of the top of the cloud deck (P$_c$) for the Phoenix models. 

The Sonora model fits show a general increase in $f_{sed}$ with later spectral type. In their model fits to 14 mid-L to mid-T brown dwarfs through the near and mid-infrared, \cite{Stephens_2009} found that earlier-type brown dwarfs generally had lower values of $f_{sed}$, while T-type objects either had high $f_{sed}$ or were cloudless. While our fits show more variation for earlier-type objects, the general trend of our results does agree with this study.

The Phoenix model fits follow a general trend of increasing grain size and increasing P$_c$, corresponding to (vertically) thinner cloud decks. These results agree with those of \cite{Brock_2021}; when they performed model fits to 16 L and T-type brown dwarfs, they also found that larger mean grain sizes and thinner cloud decks were needed to fit T dwarf spectra.

The low $f_{sed}$ and small $a_0$/P$_c$ values in the early and mid L-types point to L dwarfs having substantial cloud decks composed of fairly small particles.{ Larger $a_0$/P$_c$ values and higher $f_{sed}$ values in the early Ts suggest that these objects have clouds that are thinner and/or deeper in the atmosphere, and are composed of larger individual particles than the L-dwarf clouds.}

{As a brown dwarf cools through the L/T transition, the temperature/pressure regime at which cloud decks of different species might form falls deeper and deeper into the atmosphere. This explains the increasing trend in $f_{sed}$ and P$_c$: as we move through the spectral type sequence, the silicate cloud deck falls deeper into the atmosphere and thus has less of an impact on the observed spectrum (Caroline Morley, priv. comm.). }

{It is interesting that $f_{sed}$=8 persists through the mid T-types, since \citet{suarez22} observe that the mid-infrared silicate absorption feature at 8-11$\mu$m disappears after L8. This discrepancy is likely due the fact that the near-infrared spectra are able to probe deeper into the atmosphere of an object than mid-to-late infrared spectra can. (For example, see panel (b) of Figures 5 and 6 in \citet{mccarthy24b} where the authors plot the contribution function for a Sonora Diamondback model with T$_{eff}$ = 1100K and log($g$)=4.5. The near infrared spectrum clearly originates from a deeper level of the atmosphere than the spectrum at 8-11 $\mu$m.) It is possible that L8 is the spectral type where the silicate clouds fall too deep to be seen in the mid-infrared, but they do not fall below the probing depth of the near-infrared until the mid T-types.}

{Furthermore, \citet{cushing06} show in Figure 9 of their paper that small grain sizes (e.g. 0.1-1 $\mu$m) are needed in order to produce the 8-11 $\mu$m feature. Both the Diamondback and Phoenix fits show an increase in mean grain size through the T-types, giving another reason for the 8-11 $\mu$m feature to disappear before the silicate clouds have completely fallen below the photosphere.}

{We note that neither the Diamondback nor the Phoenix model sets include sulfide condensates. Thus, the $f_{sed}$=8 values seen in the mid T-types cannot be indicative of the formation of the thin salt clouds predicted by \cite{Morley_2012}. }

\subsection{Mixing}

\begin{figure}[htb!]
    \centering
    \gridline{\fig{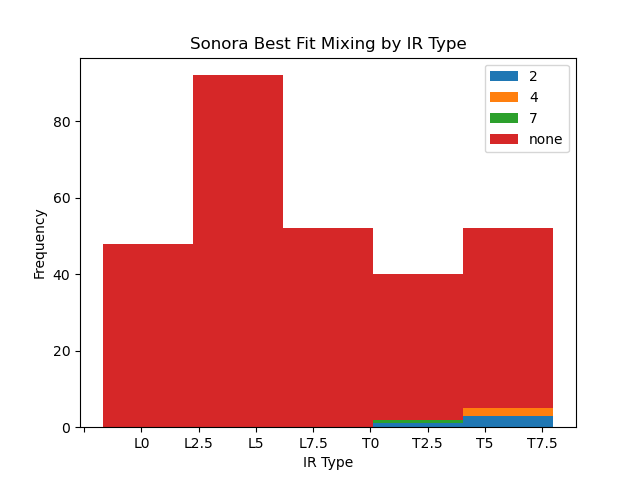}{0.5\textwidth}{(a)}
        \fig{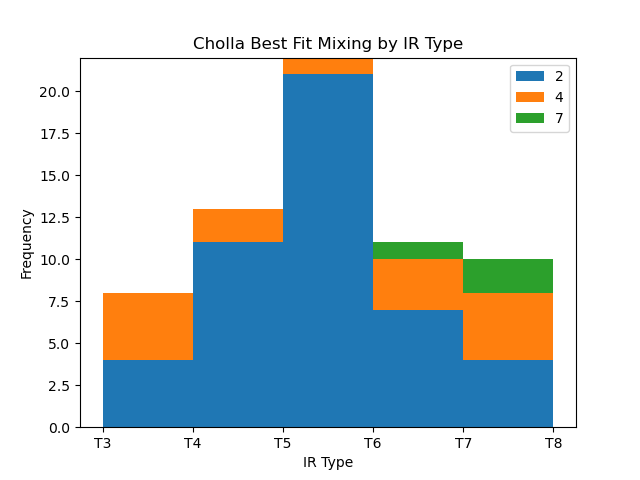}{0.5\textwidth}{(b)}}
    \gridline{\fig{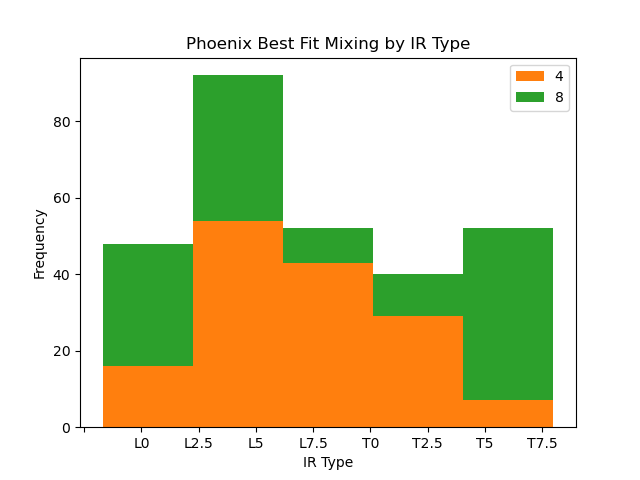}{0.5\textwidth}{(c)}}
    \caption{Histograms showing the distribution of best-fit mixing (kzz) vs infrared spectral type for the Sonora (a), Cholla only (b), and Phoenix (c) model fits. A suite of models with a full grid of cloud and mixing parameters is needed to better analyze vertical mixing.}
    \label{fig:mixing}
\end{figure}

Figure \ref{fig:mixing} gives the distribution of best-fit log $K_{zz}$. Because the Cholla model set is the only Sonora subset that accounts for disequilibrium chemistry, mixing is "none" for all of the Sonora fits except for the very few T-type objects that were best fit with Cholla models. 

In order to better investigate possible trends in mixing, we fit the objects in our sample with spectral type of T3 or later again, this time using only the Cholla models. (Note that the Cholla models only go up to 1300K, so it only made sense to fit objects past the L/T transition.) The majority of the objects in the sample were best fit with  log $K_{zz}$ = 2, although some higher values of $K_{zz}$ were found in all spectral type bins. 

The Phoenix models only have mixing value options of 4 and 8. From L0-L5, the best fit mixing values are almost evenly split. Through the L/T transition, log $K_{zz}$ = 4 takes over. After T5, log $K_{zz}$ = 8 dominates. However, the Phoenix grid used in this project did not have every pressure/$K_{zz}$ combination available for every temperature/gravity combination. Thus, it is difficult to analyze mixing on its own for the sample as a whole.

The results of \cite{Mukherjee_2022} showed that disequilibrium chemistry can notably impact the \textit{H}- and \textit{K}-bands of 1000 K atmospheres. The impact of disequilibrium chemistry on the near-infrared was progressively less obvious for atmospheres of 700 and 500 K (See Figure 2 in their paper). Based on the temperature-spectral type plots in Figure \ref{fig:tempSpecType}, we would expect to see significant disequilibrium chemistry--represented by high $K_{zz}$ values-- in the mid- and late-T spectral types. 

Retrievals by \cite{Burningham_2021} showed that disequilibrium chemistry might also be important for late-L spectral types as well. However, we do not see this trend in the Phoenix fits and the Cholla temperature grid only extends up to 1300 K, which is too cool for L-type objects. Thus, even if disequilibrium chemistry is occurring in the L-type atmospheres, it will never show up in best fits using the current Sonora models.

\cite{Mukherjee_2022} noted that mean cloud grain size is also impacted by disequilibrium chemistry. A true survey of the extent of disequilibrium chemistry in brown dwarf atmospheres requires a suite of models with a full grid that accounts for both clouds and disequilibrium chemistry. Neither the Sonora nor the Phoenix model sets have a complete cloudy, disequilibrium chemistry grid at this time. 

Furthermore, the same figure and analysis in \cite{Mukherjee_2022} also showed that the \textit{M} band provides important disequilibrium chemistry information. Extending the fits to include mid-infrared data from 3-5 $\mu$m could help to disentangle the effects of cloud grain size and disequilibrium chemistry.

\section{Interesting Families}\label{sec:interestingfamilies}

After inspecting all of the archival spectra and their model fits by-eye, we identified objects with spectral features that deviated from "typical" objects of the same spectral type. We classified these objects into families, where all the spectra in each family showed the same "interesting" spectral feature. In this section, we discuss the properties of four "interesting families" and possible reasons for their spectral oddities.

\subsection{Triangular \textit{H}-Band}

\begin{deluxetable}{ccccccc}[htb!]\label{tab:triangularHBand}
\tablecaption{Triangular \textit{H}-Band Family Members}
\tablehead{\colhead{Object} & \colhead{Spectral Type} & \colhead{2MASS \textit{J-K$_s$}} & \colhead{Abs Value R2} & \colhead{Parabolic R2} & \colhead{Sonora log($g$)}& \colhead{Phoenix log($g$)}} 
\startdata
2MASS J01415823-4633574 & L0$\gamma$, L0:VL-G & 1.73 & 0.996 & 0.926 & 5.5 & 4.74 +/- 0.25\\
2MASS J21481628+4003593 & L6, L6:FLD-G & 2.382 & 0.988 & 0.932 & 4 & 5.5\\
2MASSI J0117474-340325 & L1$\beta$, L1:INT-G & 1.689 & 0.978 & 0.908 & 5.115 +/- 0.21 & 4.5\\
SDSSp J224953.45+004404.2 & L4$\gamma$, L3:INT-G & 2.229 & 0.976 & 0.904 & 4.5 & 4.69 +/- 0.24\\
2MASSW J2244316+204343 & L6.5, L6:VL-G & 2.454 & 0.973 & 0.964 & 4 & 4\\
SDSSp J010752.33+004156.1 & L8, L5.5 & 2.115 & 0.972 & 0.958 & 4.5 & 5.5\\
2MASS J20025073-0521524 & L5$\beta$, L7 & 1.899 & 0.965 & 0.931 & 4.5 & 4.5\\
2MASSW J0030300-145033 & L7, L6:FLD-G & 1.797 & 0.962 & 0.912 & 5 & 5.5\\
2MASS J23174712-4838501 & L4, L6.5 pec (red) & 1.969 & 0.962 & 0.913 & 4.5 & 5\\
\enddata
\tablecomments{References for the spectral types shown here are given in Table \ref{tab:speXData}. Spectral type order is optical, infrared.}
\end{deluxetable}

\begin{figure}[htb!]
    \plotone{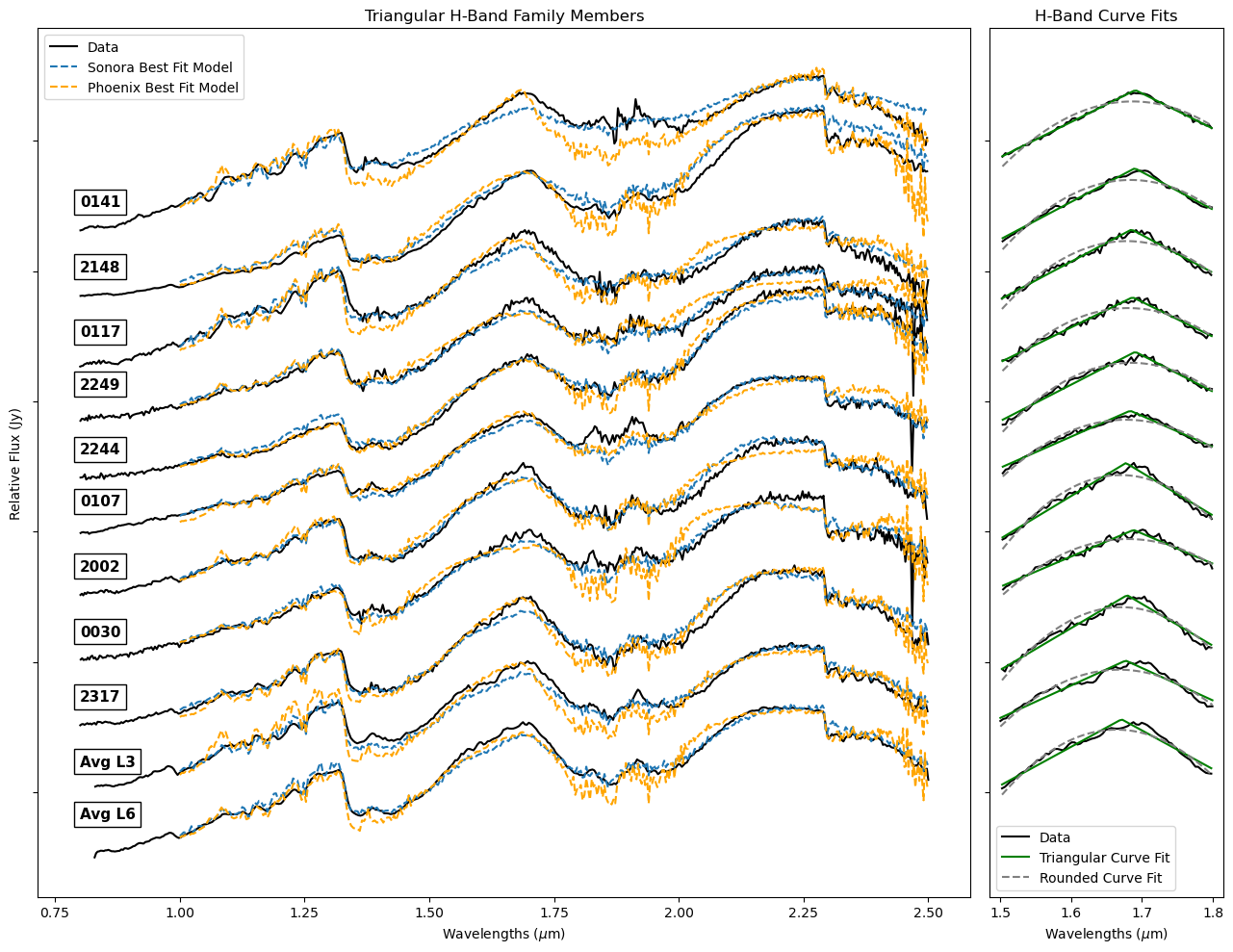}
    \caption{The SpeX spectrum (black) of each Triangular \textit{H}-band Object is shown with its best-fit Sonora (blue dashed) and Phoenix (orange dashed) models, stacked in order of decreasing adjusted $R^{2}$. The first four digits of the RA coordinates of each object are given to the left. Reference spectra made by averaging all L3 and L6 objects from the sample are included at the bottom of the plot. The Phoenix models generally do a better job at fitting the \textit{H}-band peak, while the Sonora models are generally better at fitting the \textit{K}-band. The right panel shows just the \textit{H}-band of each object (black) along with its absolute value fit (green) and parabolic fit (grey dashed).}
    \label{fig:triangularHBandFitsl}
\end{figure}

\begin{figure}[htb!]
    \plotone{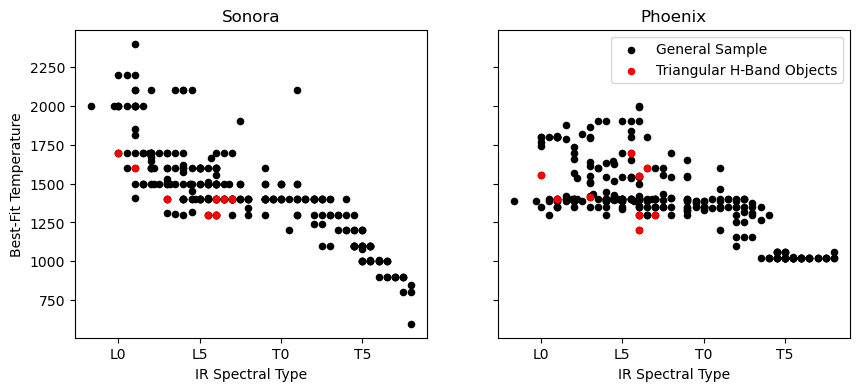} \plotone{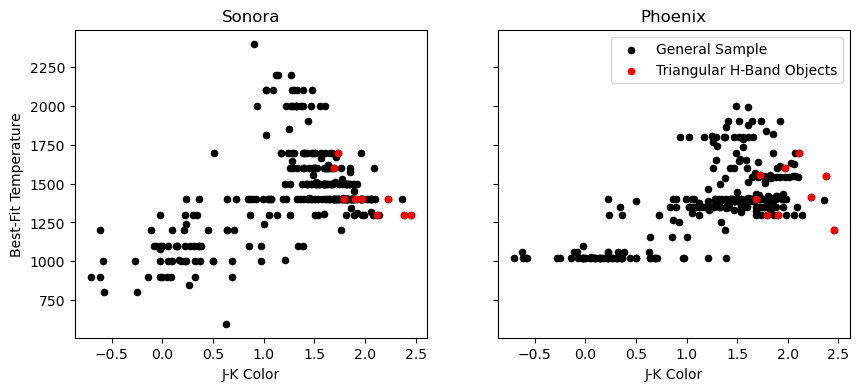}
    \caption{The objects with Triangular \textit{H}-Bands peaks are marked in red on top of plots of the Sonora (left) and Phoenix (right) model best-fit temperatures versus infrared spectral types (top) and 2MASS \textit{J-K} color (bottom). These objects are all L-types and generally fall on the cooler end for their spectral types based on their Sonora temperatures. There is no apparent pattern according to their Phoenix temperatures. Objects in this family generally have red colors for their spectral type.}
    \label{fig:triangularAndGeneralTrends}
\end{figure}

We visually identified a handful of objects in our sample with \textit{H}-band peaks at 1.7 $\mu$m that were notably sharper than the \textit{H}-band peaks of other objects of similar spectral types. All of these objects were L-dwarfs, with spectral types ranging from L0 to L6.5. 

In order to quantify peak sharpness, we used the the curve\_fit function from the scipy.optimize library to fit the H-bands (1.5-1.8 $\mu$m) of all the L dwarfs in our full sample with both a parabolic and an absolute value function. For the parabolic fits (Equation \ref{eq:parabola}), a, b, and c were left as free parameters. Likewise, the absolute value fits (Equation \ref{eq:triangle}) left a, h, and k as free parameters. We sorted all of the objects by the adjusted {coefficient of determination ($R^{2}$)} from their absolute value fits and found that the objects we had visually identified as having sharp \textit{H}-band peaks surfaced at the top of the list, along with a few other similar objects that we had not originally flagged. In the end, we kept the nine objects with the highest adjusted $R^{2}$.

\begin{equation}
\label{eq:parabola}
    y = ax^2+bx+c
\end{equation}

\begin{equation}
\label{eq:triangle}
    y = a|x-h|+k
\end{equation}

We refer to this group of deviant objects as the "Triangular \textit{H}-Band" family. Table \ref{tab:triangularHBand} lists the object names, spectral types, \textit{J-K} colors, best-fit log($g$) values, and adjusted $R^2$ values from both the absolute value and parabolic fits. Figure \ref{fig:triangularHBandFitsl} shows the SpeX spectra of these objects along with their Phoenix and Sonora model fits and their \textit{H}-band curve fits. We also calculated the average spectra of all 16 L3-type and all 28 L6-type objects and include these two average spectra in Figure \ref{fig:triangularHBandFitsl} for reference. Figure \ref{fig:triangularAndGeneralTrends} gives the locations of these family members in temperature-spectral type and temperature-color space, respectively. 

Both model sets were able to match the shape of the spectra from 0.8-1.7 $\mu$m well. The models and data do not always agree past 1.7 $\mu$m, although the Phoenix models do slightly better at fitting the \textit{H}-band peak while the Sonora models slightly better fit the shape of the \textit{K}-band.

This is the least mysterious of the four families. A triangular \textit{H}-band profile has long been seen as a flag of low gravity and youth.  \citet{Kirkpatrick_2006} suggest that the lower pressures of low-gravity atmospheres lead to a reduction in collision induced absorption (CIA) of H$_2$ compared to higher-gravity atmospheres. They believed this reduction in CIA to be the cause of the triangular \textit{H}-band profile.

They also note the redder-than-average \textit{J-K} color of 2MASS J0141-4633, one of the members of this family. All of the objects we identified as being a part of the Triangular \textit{H}-Band Family have red colors, as shown in Figure \ref{fig:triangularAndGeneralTrends}. Other works have also linked red colors and youth \citep{Kirkpatrick_2008,Cruz_2009,Faherty_2012,Faherty_2016,Schneider_2023}. \citet{Barman_2011} suggest that the atmospheres of low-gravity objects are able to maintain thicker cloud decks than higher-gravity objects. The thicker cloud decks and lower CIA likely work together to produce the redder colors.

Even though they all appear to be low-gravity objects, only one object had a best-fit log($g$) \textless4.5 based on the Phoenix models, while only two had log($g$) \textless4.5 according to the Sonora model fits. We believe this is an indication that surface gravity is poorly constrained in model fits to the entire near-infrared, not a reason to reject the low-gravity classification. 

Only three of the objects we identified as members of this family have low-gravity $\gamma$ spectral-type markers listed in the SPL, even though they all have sharper-than-average \textit{H}-bands. Two more objects have intermediate-gravity $\beta$ markers and one has a "peculiar, red" marker. Two of the remaining objects have been classified as field-gravity. One of these (2MASS J0030-1450) does have the weakest triangular peak in the sample, but the other (2MASS J2148+4003) has the second-highest $R^2$. 

\subsection{Plateaued \textit{H}-Band}

\begin{deluxetable}{cccccc}[htb!] \label{tab:plateauedH}
\tablecaption{Plateaued \textit{H}-Band Family Members}
\tablehead{\colhead{Object} & \colhead{Spectral Type} & \colhead{2MASS \textit{J-K}$_s$} & \colhead{Trapezoid R2} & \colhead{Parabolic R2} & \colhead{Abs Value R2}}
\startdata
    SDSS J120747.17+024424.8 & L8, T0 & 1.594 & 0.9860 & 0.9625 & 0.8470\\
    SDSSp J042348.57-041403.5 & L7.5, T0 & 1.536 & 0.9857 & 0.9693 & 0.8761\\
    2MASS J09121469+1459396 & L8, T0 & 1.469 & 0.9799 & 0.9676 & 0.8538\\
    SDSS J205235.31-160929.8 & T1: & 1.673 & 0.9795 & 0.9586 & 0.8102\\
    SDSS J015141.69+124429.6 & T1 & 1.383 & 0.9787 & 0.9603 & 0.8129\\
\enddata
\tablecomments{References for the spectral types shown here are given in Table \ref{tab:speXData}. Spectral type order is optical, infrared.}
\end{deluxetable}

\begin{figure}[htb!]
    \plotone{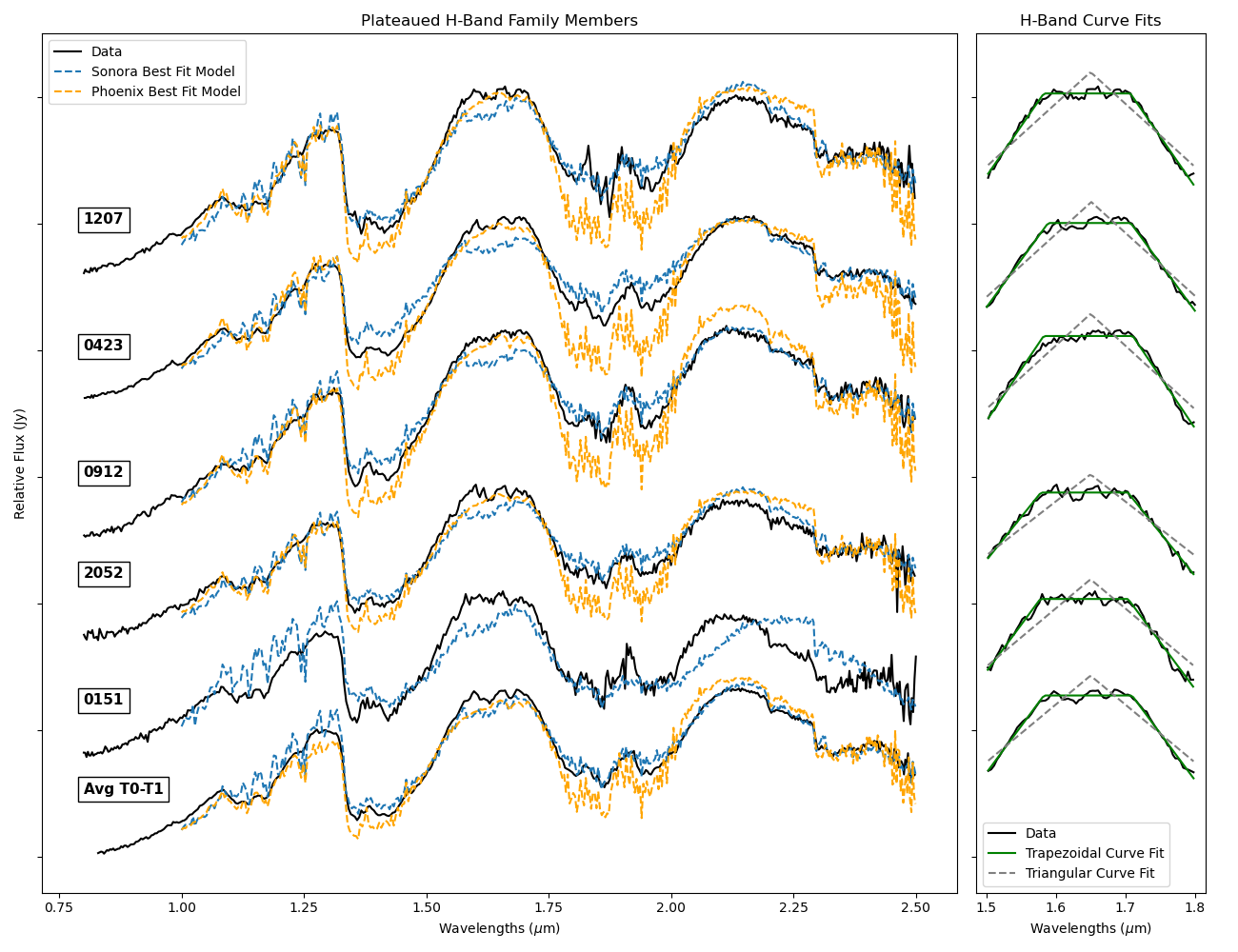}
    \caption{The SpeX spectrum (black) of each plateaued \textit{H}-band object is shown with its best-fit Sonora (blue dashed) and Phoenix (orange dashed) models, stacked in order of decreasing adjusted $R^{2}$. The first four digits of the RA coordinates of each object are given to the left. A reference spectra made by averaging all T0 and T1 objects from the sample is included at the bottom of the plot. The Phoenix models generally do a better job at fitting the \textit{H}-band peak, although both models give generally poor fits through the \textit{H} and \textit{K} bands. The Phoenix fit for 0151 completely fails to converge to a meaningful answer, and thus is not included in the plot. The right panel shows just the \textit{H}-band of each object (black) along with its trapezoidal fit (green) and absolute value fit (grey dashed).}
    \label{fig:plateauedHBandFitsl}
\end{figure}

\begin{figure}[htb!]
    \plotone{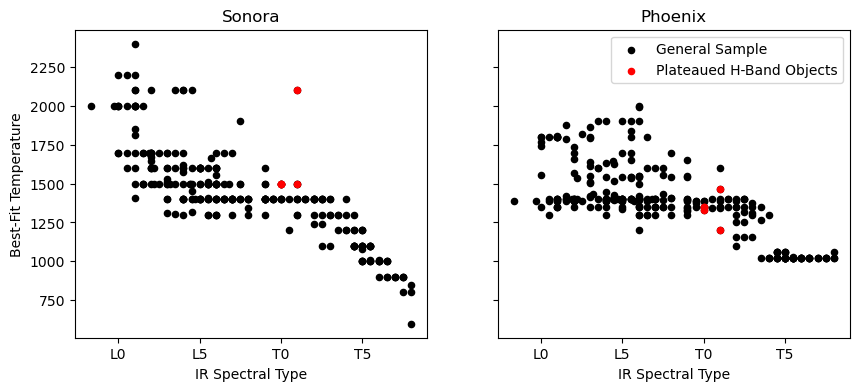} \plotone{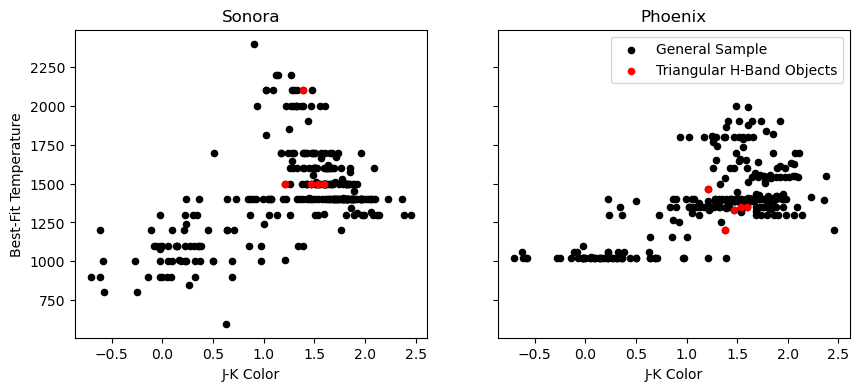}
    \caption{The objects with Plateaued \textit{H}-Bands peaks are marked in red on top of plots of the Sonora (left) and Phoenix (right) model best-fit temperatures versus infrared spectral types (top) and 2MASS \textit{J-K} color (bottom). These objects are all early T-types and fall on the warmer end for their spectral types based on their Sonora temperatures. There is no apparent pattern according to their Phoenix temperatures. Objects in this family have fairly average colors for their spectral type.}
    \label{fig:plateauAndGeneralTrends}
\end{figure}

The second family of objects consists of L/T transition types with flat-topped, plateau-like \textit{H}-band peaks. We call this group the "Plateaued \textit{H}-Band Family."

To quantify the "flatness" of the objects in our sample, we fit the \textit{H}-bands of all of the L/T transition-type objects with a trapezoidal function (see Equation \ref{eq:trapezoid}) with four free parameters: a, h, k, and "side," which determines the width of the trapezoid top. We sorted the objects by adjusted $R^{2}$. Out of the top nine objects, we kept five that we visually confirmed to be abnormally "flat." The other four were flat enough to be well-fit by the trapezoidal curve, but were thrown out due to their \textit{J}-band peaks that slope up in the direction of bluer wavelengths and jagged \textit{H}-band plateau surfaces that better match the next two families of objects. 

\begin{equation}
\label{eq:trapezoid}
    y = 
    \begin{cases}
    a\left|x-h+side\right|+k  &\text{if } x \leq h-side,\\
    k  &\text{if } x \in [h-side, h+side),\\
    a\left|x-h-side\right|+k  &\text{if } x \geq h+side.
    \end{cases}
\end{equation}

The list of objects and their model fits are given in Table \ref{tab:plateauedH} and Figure \ref{fig:plateauedHBandFitsl}, respectively. We also include a reference spectrum made by taking the average of all 20 T0- and T1-type objects in our sample. In this case, the average spectrum looks like it belongs in this family of objects. {However, this is deceptive as only 5/20 of the T0- and T1-type objects in our sample display true flat-topped H-bands. The L/T transition includes a wide range of \textit{H}-band peak shapes that average out to a flat top. For example, the next two families are also made up of L/T transition objects, but their \textit{H}-band peaks look notably different from this family. }

The model fits to this family are not as clean as the fits to the Triangular \textit{H}-Band family members. In fact, the Phoenix fit to SDSS J0151+1244 fails entirely, likely due to this object's abnormally low \textit{K}-band flux. With the exception of 0151, the Phoenix models do a better job of matching the \textit{H}-band height than the Sonora models. Neither model set is perfectly able to match the flatness of the peaks, though: the Phoenix models are too rounded while the Sonora models include too much absorption around 1.6 $\mu$m. Both model sets tend to underfit the \textit{H}-band and overfit the \textit{K}-band. 

A literature search reveals that J1207+0244 is a suspected binary (L6.5+T2.5; \cite{Burg10}), while J0423-0414 (L6.5+T2; \cite{dupuy12}), J0912+1459 (L8.5+L7.5; \cite{dupuy12}), and J2052-1609 (L8.5+T1.5; \cite{dupuy12}) are confirmed binaries. We found no studies where J0151+1244 had been flagged as a potential binary. 
\begin{deluxetable}{cccccc}[htb!] \label{tab:binaryPlateauedH}
\tablecaption{Plateaued \textit{H}-Band Binary Fit Results}
\tablehead{\colhead{Object} & \colhead{Temp} & \colhead{log($g$)} & \colhead{fsed} & \colhead{log(kzz)} & \colhead{metallicity}}
\startdata
    SDSS J120747.17+024424.8 & 1400/1400 & 5/5 & 2/8 & none/none & 0.5/0\\
    SDSSp J042348.57-041403.5 & 1400/1500 & 5/5.5 & 8/3 & none/none & 0.5/0\\
    2MASS J09121469+1459396 & 1000/1400 & 5.5/5 & 4/4 & none/none & -0.5/0\\
    SDSS J205235.31-160929.8 & 1250/1500 & 3.5/5.5 & none/8 & 0.7/none & 0/0\\
    SDSS J015141.69+124429.6 & 1200/1400 & 5.5/5 & 2/8 & none/none & -0.5/0\\
\enddata
\end{deluxetable}

\begin{figure}[htb!]
    \plotone{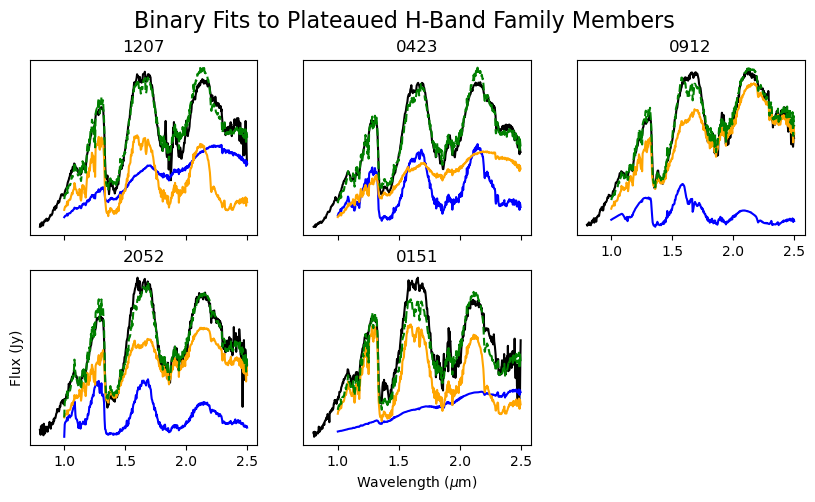}
    \caption{Plots showing binary model fits to the five members of the Plateaued \textit{H}-Band Family. The models are all Sonora models. The data and composite models are plotted in black and green, respectively, The orange and blue curves give the individual components that were summed together to generate the composite model. The parameters of the components are given above in Table \ref{tab:binaryPlateauedH}}
    \label{fig:plateauBinaries}
\end{figure}

{Because 3/5 of the members of this family are confirmed binaries, we hypothesize that the plateaued H-band morphology is indicative of binarity.} We performed binary fits to each of these family members {to see if composite models could completely explain the flat H-bands.} The parameters of the best-fit composite models are given in Table \ref{tab:binaryPlateauedH}, while Figure \ref{fig:plateauBinaries} displays the fits. The binary models generally improved either the \textit{H} or \textit{K}-band fits, never both. {Thus, binarity alone might not be the only explanation for the unique spectral characteristics of this family.}

Inspection of the best binary fit parameters reveals no obvious trend. The two component models for each object have different combinations of log($g$), f$_{sed}$, metallicity, and in one case even different log($K_{zz}$) values when one component model came from the Cholla set. The mismatch between the models and data and strange combination of composite model component parameters suggest that some physical process is taking place in the atmospheres of these objects that is not completely represented in current theory.

This struggle of both the binary and single models to simultaneously fit the \textit{H}-band (containing the CH${_4}$ feature at 1.6 $\mu$m) and the \textit{K}-band (containing important CH${_4}$ and CO bands) could potentially be a result of disequilibrium carbon chemistry. The best Sonora fits to these objects were all Diamondback models with the exception of 0151, which matched with a Bobcat. Neither the Bobcats nor the Diamondbacks account for disequilibrium chemistry. The Phoenix models do allow for some vertical mixing, but the $K_{zz}$ grid only has options of log $K_{zz}$ =4 or 8. Thus, neither model set is ideal for studies of disequilibrium chemistry. 

\begin{figure}[htb!]
    \centering    
    \plotone{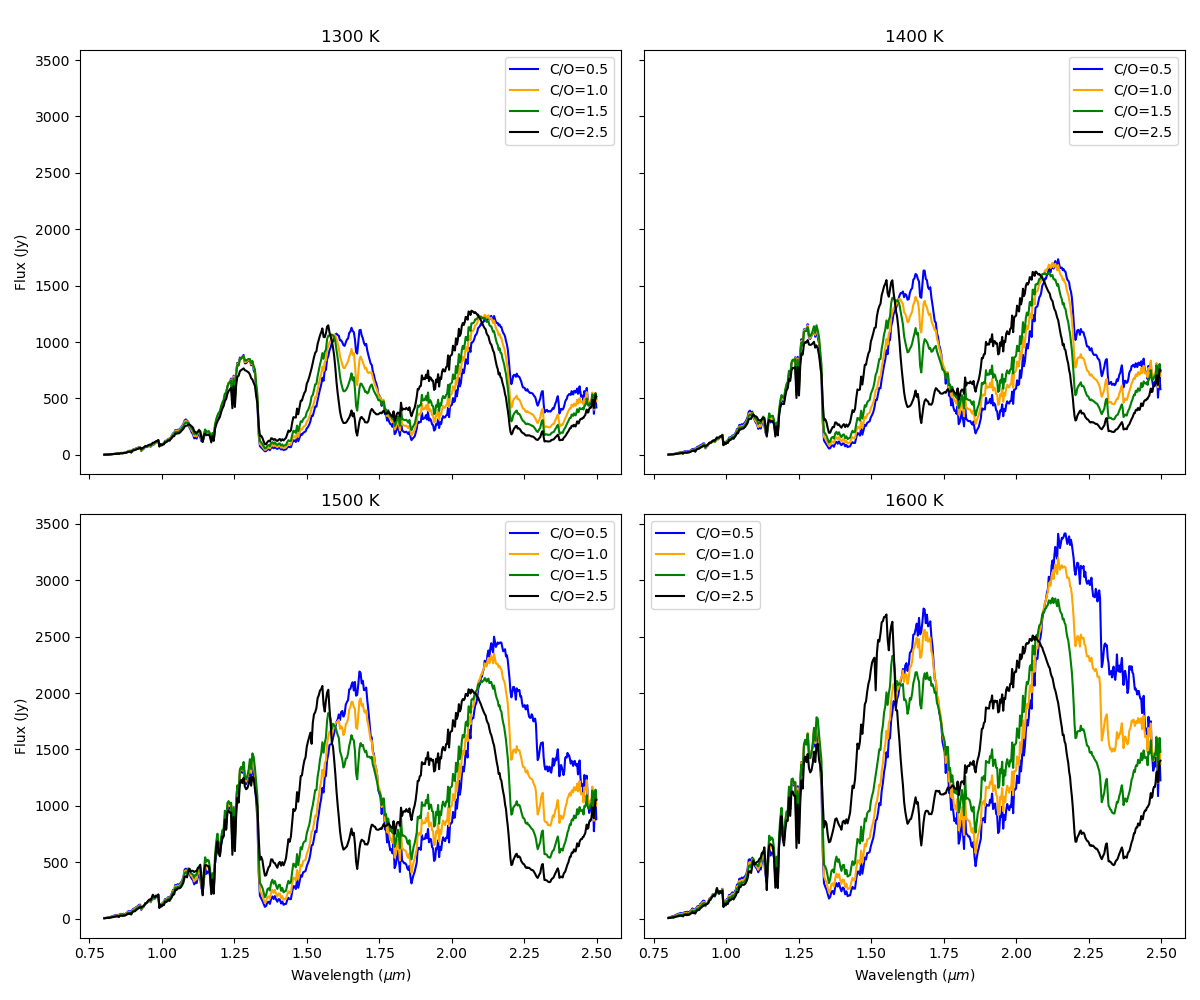}
    \caption{Plots of Sonora Elf Owl models \citep{Mukherjee_2024_elfowl} with L/T transition-type temperatures and varied C/O ratios. C/O ratio alone is not enough to explain the flat shapes of the Plateaued H-Band family members, since even the flattest models plotted here still show too much absorption around 1.6/1.65 $\mu$m.}
    \label{fig:co}
\end{figure}

We also investigate C/O ratio as a possible cause of the plateaued H-band shape. Because neither the Diamondback nor the Cholla models allow for a non-solar C/O ratio, we use the Sonora Elf Owl models \citep{Mukherjee_2024_elfowl,ZenodoElfOwl} in Figure \ref{fig:co} to visually identify how changing the C/O ratio impacts the near-infrared spectrum of an L/T transition-type object. All of the models plotted in this figure have log($g$) = 5.0 and solar metallicity. 

The Elf Owl models do predict that the C/O ratio can impact the flattness of the H-band. For example, a C/O ratio of 0.5 gives the flattest H-band for the models with T$_{eff}$=1300 K, while a C/O ratio of 1.5 gives the best plateau for the 1600 K models. However, these models all still display more jagged absorption features around 1.6/1.65 $\mu$m than are seen in the members of the Plateau family. Thus, a nonsolar C/O ratio by itself does not fully explain the H-band plateaus.

Future work could involve fitting these objects with a retrieval code or the open-source code PICASO 3.0 \citep{Mukherjee_2022,Mukherjee_2023} in order to be able to simultaneously fine-tune the extent of disequilibrium chemistry, cloud parameters, {C/O ratio,} and metallicity. It would be beneficial to include longer wavelengths in this proposed analysis to help to disentangle the somewhat degenerate effects of the above parameters.

\subsection{Double-Peaked \textit{H} Band}

\begin{deluxetable}{ccccc}[htb!]\label{tab:doublePeak}
\tablecaption{Double-Peak Family Members}
\tablehead{\colhead{Object} & \colhead{Spectral Type} & \colhead{2MASS \textit{J-K}$_s$} & \colhead{Binarity} & \colhead{Binarity Ref}} 
\startdata
    SDSS J035104.37+481046.8 & T1: & 1.47 & S:L6.5/T5 & 1\\
    2MASS J05185995-2828372 & L7,T1p & 1.816 & C:L6/T4 & 2\\
    SDSS J080531.84+481233.0 & L4,L9: & 1.29 & C:L4/T5 & 2\\
    SDSS J151114.66+060742.9 & T2 & 1.472 & C:L5/T5 & 3\\
    2MASSI J1711457+223204 & L6.5,L9:: & 2.362 & S:L5/T5.5 & 1\\
    SDSS J151603.03+025928.9 & T0: & 1.797 & S:L7.5/T2.5 & 1\\
    SDSS J024749.90-163112.6 & T2 & 1.57 & S:T0/T7 & 1\\
\enddata
\tablecomments{The "binarity" column gives the two component spectral types proposed in the literature. "S" means that the object is a suspected spectral binary, while "C" indicates a confirmed binary. Spectral type references are given in Table \ref{tab:speXData}. Spectral type order is optical,infrared.}
\tablerefs{(1) \cite{Burg10}; (2) \cite{dupuy12}; (3) \cite{gag15}}
\end{deluxetable}

\begin{figure}[htb!]
    \plotone{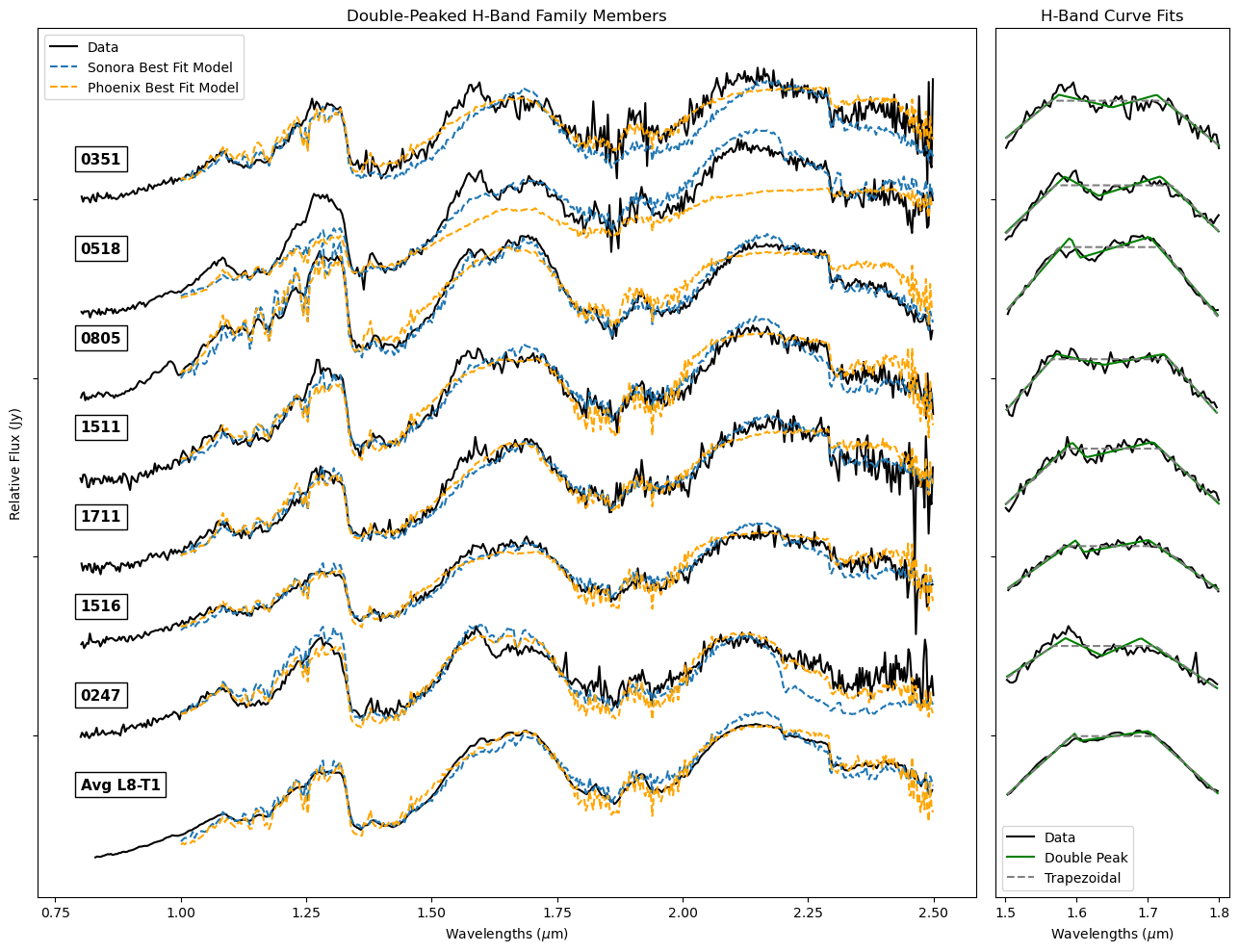}
    \caption{The SpeX spectrum (black) of each Double-Peaked \textit{H}-band object is shown with its best-fit Sonora (blue dashed) and Phoenix (orange dashed) models. The first four digits of the RA coordinates of each object are given to the left. A reference spectrum made by averaging all L8-T1 objects from the sample is included at the bottom of the plot. The right panel shows just the \textit{H}-band of each object (black) along with its double-peaked fit (green) and trapezoidal fit (grey dashed).}
    \label{fig:doublePeakHBandFitsl}
\end{figure}

\begin{figure}[htb!]
    \plotone{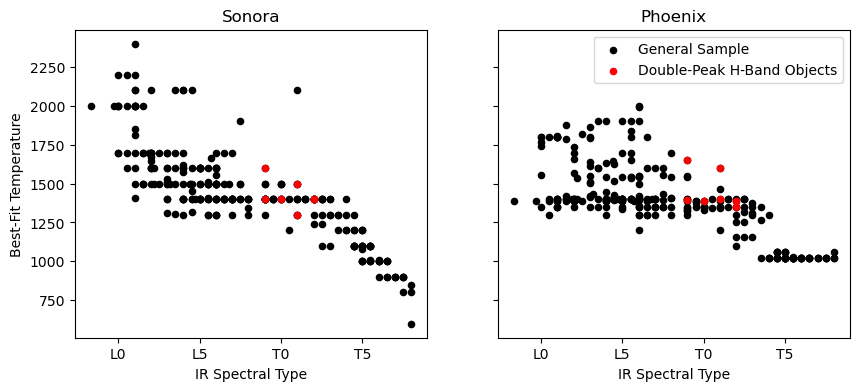}  \plotone{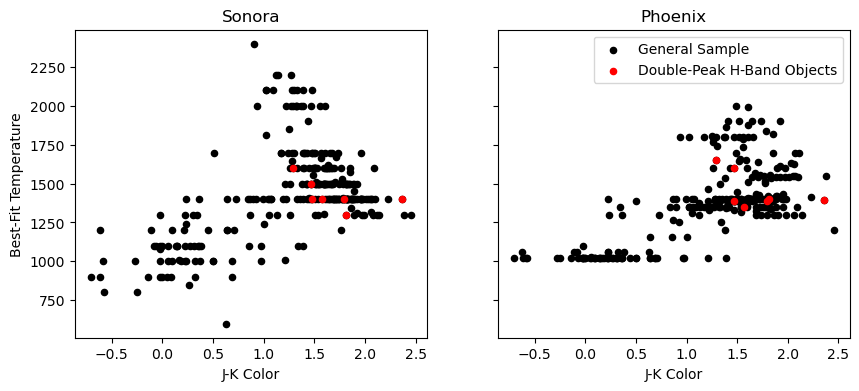}
    \caption{The objects with Double-Peaked \textit{H}-Bands are marked in red on top of plots of the Sonora (left) and Phoenix (right) model best-fit temperatures versus infrared spectral types (top) and 2MASS \textit{J-K} color (bottom). All members of this family fall in the L/T transition. There is not a consistent trend in \textit{J-K} color for this family.}
    \label{fig:doublePeakAndGeneralTrends}
\end{figure}

Objects in this family have a jagged, double-spiked top of the \textit{H}-band peak around 1.6 $\mu$m. We first identified the core family members visually. Then, we fit each of the core members with an adapted version of our fitting code that uses all of the SpeX data files as the models, allowing us to quantitatively group together similar objects. In total, we ended up with seven members of this family.{ The family members span types L8-T1, and make up $\sim$18\% of the objects in this spectral range.}

The names, spectral types, \textit{J-K} colors, and binarity status of these objects are given in Table \ref{tab:doublePeak}. Plots of the SpeX spectra and model fits of the family members are shown in Figure \ref{fig:doublePeakHBandFitsl}. 

We also used curve\_fit to fit the \textit{H}-bands of the L/T transition objects with the double-peak shape given in Equation \ref{eq:double}. Sorting by adjusted $R^2$ was not an effective way to define this family, because many different peak shapes are well-fit by Equation \ref{eq:double} thanks to its six free parameters. We do include the best-fit double-peak curve for each family member in Figure \ref{fig:doublePeakHBandFitsl} to help with visualization.

\begin{equation}
\label{eq:double}
    y = 
    \begin{cases}
    a\left|x-h+side\right|+k  &\text{if } x < h-side,\\
    \frac{ypivot-k}{xpivot-h+side}(x-h+side)+k  &\text{if } x \in [h-side, xpivot],\\
    \frac{k-ypivot}{h+side-xpivot}(x-xpivot)+ypivot  &\text{if } x \in (xpivot, h+side],\\
    a\left|x-h-side\right|+k  &\text{if } x > h+side.
    \end{cases}
\end{equation}

There is a known CH$_4$ absorption feature at 1.6 $\mu$m seen in T-dwarfs that could account for the double-peaked shape. However, these are all L/T transition objects and thus are not yet expected to have much CH$_4$ in their upper atmospheres. Furthermore, none of these objects show much of an absorption feature at 2.2 $\mu$m, the other region in the near-IR where CH$_4$ is visible for T-dwarfs.

{We reject nonsolar C/O ratio as a likely cause for the morphology of this family because none of the C/O ratio/T$_{eff}$ combinations plotted in Figure \ref{fig:co} are able to recreate the double-peak shape.}

The objects in this family also have \textit{J}-band peaks with slopes angled up in the direction of bluer wavelengths. Typical objects in the L/T transition have \textit{J}-band peaks that are either angled up towards the red side of the wavelength sequence or are flat-topped. 

A literature search revealed that three of these family members are confirmed binaries, while the other four have been flagged as candidate spectral binaries (Table \ref{tab:doublePeak}). In each case except for J0247, the primary is an L-type object while the secondary is a T-type. J0247 is suspected to have a T0 primary. The variety in depth of the troughs in-between the double peaks of the various family members could be explained by slight differences in component member spectral types. For example, J1516 is suspected to have the latest L-type primary and earliest T-type secondary of the sample, and also shows the weakest double peak. In all cases but J1516, the component spectral types for each object in this family differ more significantly than the components of the Plateaued \textit{H}-Band Family members.  

\begin{figure}[htb!]
    \plotone{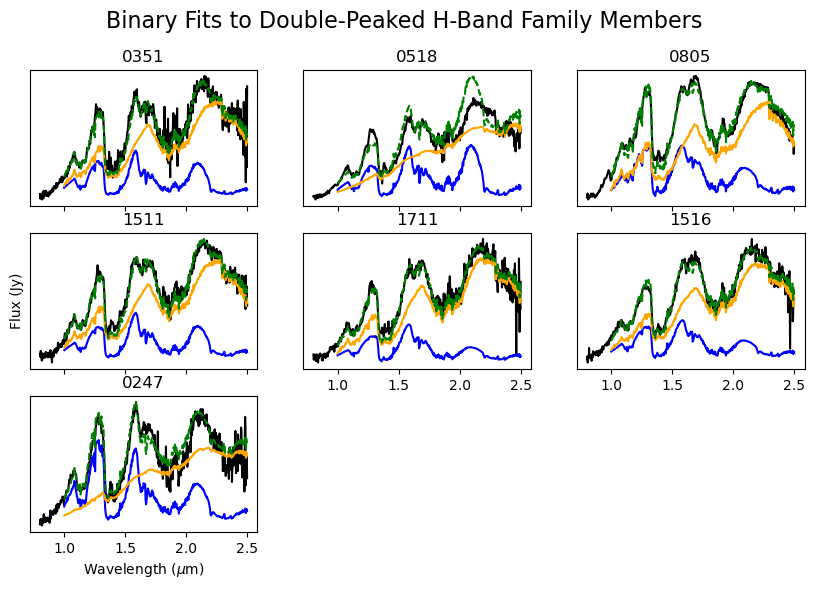}
    \caption{Plots showing binary model fits to the six members of the Double-Peaked \textit{H}-Band Family. The models are all Sonora models. The data and composite models are plotted in black and green, respectively, The orange and blue curves give the individual components that were summed together to generate the composite model. The parameters of the components are given above in Table \ref{tab:binaryDoubles}.}
    \label{fig:doublePeakBinaries}
\end{figure}

We performed binary fits (Figure \ref{fig:doublePeakBinaries}) to each of the seven family members and found that all of the objects were best-fit by binary models except for J0518, which was also very poorly fit by single models. Each primary component had a temperature consistent with the L/T transition, while the secondary component temperature was more in line with a mid T-type dwarf. The exception to this was J0247, which had two cooler component temperatures more in line with its suspected component spectral types of T0/T7. We conclude that membership of the Double-Peaked \textit{H}-Band family is an indication of binarity.

\begin{deluxetable}{cccccc}[htb!] \label{tab:binaryDoubles}
\tablecaption{Double-Peaked \textit{H}-Band Binary Fit Results}
\tablehead{\colhead{Object} & \colhead{Temp} & \colhead{log($g$)} & \colhead{fsed} & \colhead{log(kzz)} & \colhead{metallicity}}
\startdata
    SDSS J035104.37+481046.8 & 1100/1400 & 5/4 & 8/4 & none/none & 0/0.5\\
    2MASS J05185995-2828372 & 1000/1200 & 5/5.5 & 4/2 & none/none & 0/-0.5\\
    SDSS J080531.84+481233.0 & 1300/1500 & 5.5/4 & 8/4 & none/none & -0.5/0.5\\
    SDSS J151114.66+060742.9 & 1100/1400 & 5.5/4 & 8/4 & none/none & 0/0.5\\
    2MASSI J1711457+223204 & 1000/1300 & 5.5/3.5 & 4/4 & none/none & -0.5/0\\
    SDSS J151603.03+025928.9 & 1100/1400 & 5.5/4.5 & 4/4 & none/none & -0.5/0.5\\
    SDSS J024749.90-163112.6 & 1100/1200 & 4.5/4.5 & 8/1 & none/none & -0.5/-0.5\\
\enddata
\end{deluxetable}

\subsection{Blue Early Ts} \label{subsec:overluminousT}

\begin{figure}[htb!]  \plotone{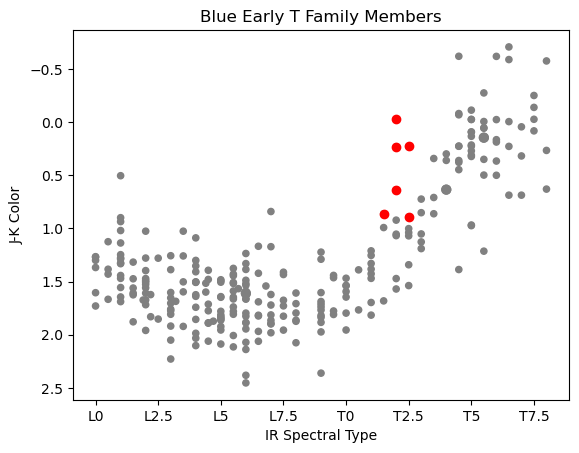}
    \caption{2MASS \textit{J-K} color is plotted against infrared spectral type. The blue early T objects are marked in red on top of the general sample (grey). The members of this family are all T1 or T2s and are blue for their spectral types.}
    \label{fig:tooHotAndGeneralTrends}
\end{figure}

We identified six early T-type objects with very blue \textit{J-K} colors for their spectral types. These objects are marked in red in Figure \ref{fig:tooHotAndGeneralTrends}. Visual comparison of the spectra of these objects to other early T-type spectra shows a difference in relative \textit{J} and \textit{K} bands; the members of this "Blue Early T" family are brighter in \textit{J}. {The family members have spectral types ranging from T1-T2.5, and make up $\sim$25\% of the objects in this spectral range.}

\begin{deluxetable}{ccccc}[htb!]\label{tab:tooHotTs}
\tablecaption{Blue Early T Family Members}
\tablehead{\colhead{Object} & \colhead{Spectral Type} & \colhead{2MASS \textit{J-K}$_s$} & \colhead{Binarity} & \colhead{Binarity Ref}} 
\startdata
    SIMP J013656.57+093347.3 & T2, T2.5 & 0.893 & -- &\\
    SDSS J090900.73+652527.2 & T1, T1.5 & 0.863 & ST1.5/T2.5 & 1\\
    2MASS J11220826-3512363 & T2 & 0.636 & -- & --\\
    SDSS J104829.21+091937.8 & T2.5 & 0.229 & -- & --\\
    SDSS J143553.25+112948.6 & T2: & 0.231 & SL7.5/T6 & 1\\
    SDSS J011912.22+240331.6 & T2 & -0.025 & ST0/T4 & 1\\
\enddata
\tablerefs{(1) \cite{Burg10} }
\tablecomments{References for the single spectral types shown here are given in Table \ref{tab:speXData}.}
\end{deluxetable}

\begin{figure}[htb!]
    \plotone{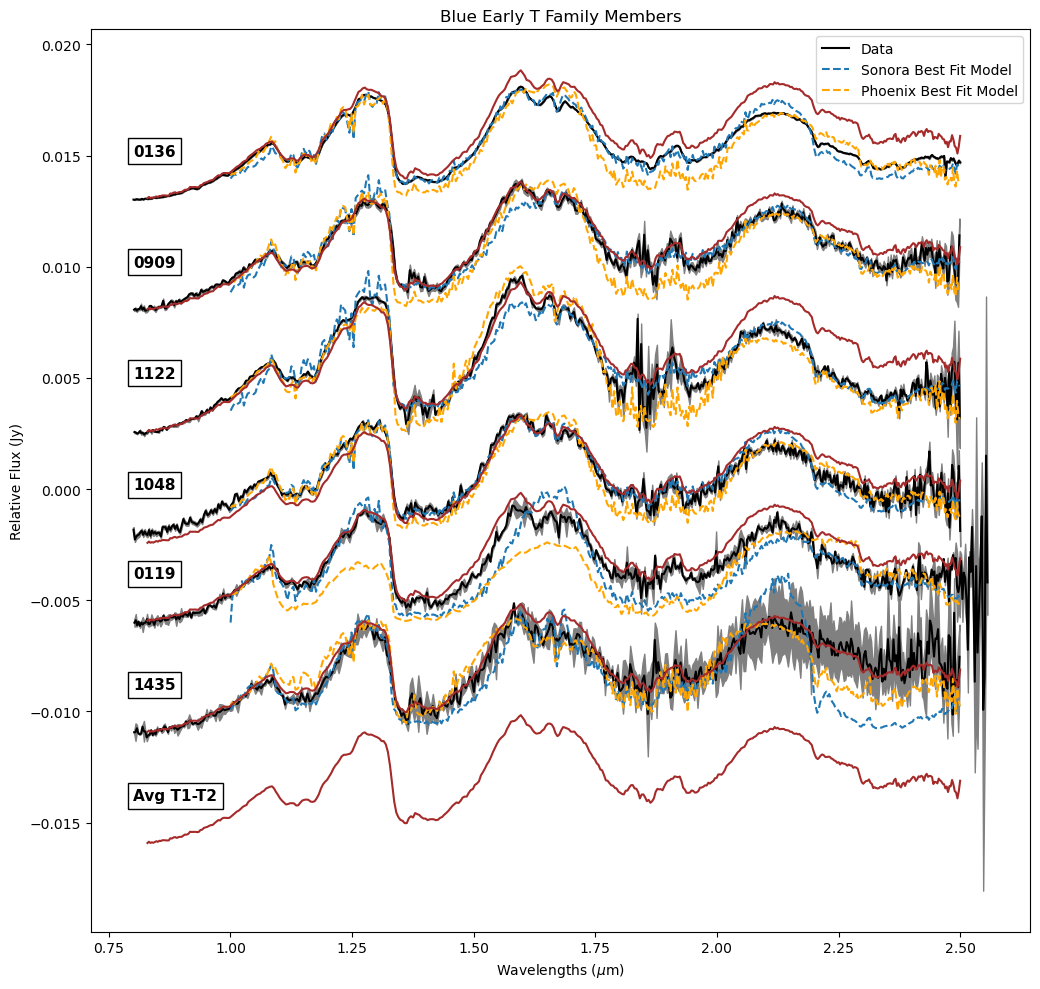}
    \caption{The SpeX spectrum (black) of each blue early T object is shown with its uncertainty (grey) and best-fit Sonora (blue dashed) and Phoenix (orange dashed) models. The first four digits of the RA coordinates of each object are given to the left. A reference spectrum made by averaging all T1-T2 objects from the sample is included at the bottom of the plot. The reference spectrum is plotted on top of each object, roughly scaled so \textit{J}-band fluxes match. Note that all members of this family are overluminous in \textit{J}-band relative to their \textit{K}-band peaks. J1435 is the exception, but its \textit{K}-band has very large uncertainty. This object was included in the family because of its very blue 2MASS \textit{J-K} color.}
    \label{fig:tooHotTFitsl}
\end{figure}

Table \ref{tab:tooHotTs} lists the family members along with their spectral types, \textit{J-K} colors, and binarity status. Each family member's spectrum as well as its best Sonora and Phoenix model fits are plotted in Figure \ref{fig:tooHotTFitsl}. A reference spectrum was created by taking the average of all T1 and T2-type objects from the general sample. The reference spectrum is plotted in brown on top of the six family members to show that each family member is overluminous in \textit{J}-band relative to \textit{K}-band, compared to more typical early T-types. The exception to this is J1435. 

J1435 was classified into the Blue Early T family based on its very blue 2MASS \textit{J-K} color, but its relative \textit{J}, \textit{H}, and \textit{K}-band peak heights match the reference spectrum well. The SpeX data for this object has notably low signal-to-noise. The \textit{H}-band uncertainties in particular are high enough that the \textit{K}-band peak could indeed be dim enough to explain the blue color and better match the other members of this family. We chose to keep this object in the family on the basis of its color, but follow-up with higher signal-to-noise data would be useful to confirm family membership. 

\begin{deluxetable}{cccccc}[htb!] \label{tab:binaryTooHots}
\tablecaption{Blue Early T Binary Fit Results}
\tablehead{\colhead{Object} & \colhead{Temp} & \colhead{log($g$)} & \colhead{fsed} & \colhead{log(kzz)} & \colhead{metallicity}}
\startdata
    SIMP J013656.57+093347.3 & 1200/1300 & 5.5/4 & 8/3 & none/none & -0.5/0\\
    SDSS J090900.73+652527.2 & 1200/1300 & 5.5/3.5 & 8/3 & none/none & -0.5/0\\
    2MASS J11220826-3512363 & 1300/1300  & 3.5/5.5  & nc/8  & 7/none &  0/-0.5\\
    SDSS J104829.21+091937.8 & 1100/1200 & 3.5/5.5 & 4/8 & none/none & 0.5/-0.5\\
    SDSS J143553.25+112948.6 & 900/1300 & 5.5/4 & nc/2 & 7/none & 0/-0.5\\
    SDSS J011912.22+240331.6 & 1300/1000 & 4/4.5 & nc/1 & 4/none & 0/0.5\\
\enddata
\end{deluxetable}

\begin{figure}[htb!]
    \plotone{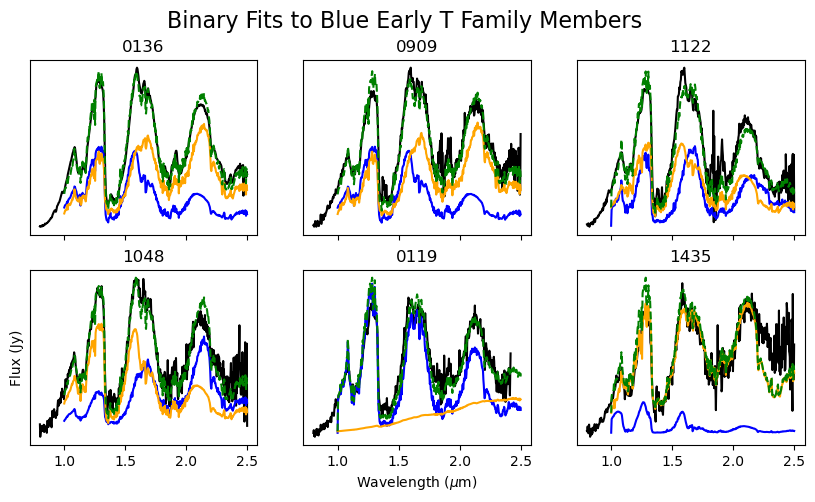}
    \caption{Plots showing the best binary model fits to the five members of the Blue Early T Family. The models are all Sonora models. The data and composite models are plotted in black and green, respectively, The orange and blue curves give the individual components that were summed together to generate the composite model. The parameters of the two component models are given in Table \ref{tab:binaryTooHots}.}
    \label{fig:overluminousBinaries}
\end{figure}

One possible explanation is binarity. We found that, for five of the six objects, binary model fits (Figure \ref{fig:overluminousBinaries}) were better able to simultaneously match the relative \textit{J}, \textit{H}, and \textit{K} peak heights than the best single model fits. The only object that did not show improvement was {J1048.} 

Each best binary fit to these family members was to a composite model made of a primary with a temperature consistent with an L/T transition spectral type (1200-1300K) and a second object with a temperature consistent with a mid-to-late T-dwarf (1200-900K). However, even the binary fits were still not perfectly able to explain the overluminosity in \textit{J}, suggesting that there is more to the story than just binarity.

A second possible explanation is that these objects are not binary but instead are experiencing inhomogeneous cloud-clearing \citep{Marley_2010,Burgasser_2002}. These objects might have initially had thick cloud decks that blocked light from escaping deeper levels of the atmosphere. Then, as the objects cooled through the L/T transition these clouds could have begun to clear away and develop patchy holes. Light escaping through those holes would have originated from deeper, hotter levels of the atmosphere than light emitted from the level of the cloud deck. The resulting spectrum in this scenario would show an increase of flux at bluer wavelengths compared to a cloudy object of the same spectral type. Furthermore, because the abundances of different molecular species change with the pressure and temperature of a brown dwarf atmosphere, flux emitted from two different temperature/pressure levels of the same object would have a mixture of spectral features from both levels. This mixing of spectral features could cause the object to be falsely flagged as a binary. Such an object would also likely be variable due to the inhomogeneous distribution of clouds \citep{Radigan_2014}.

This theory lines up with the overluminosity of the \textit{J}-bands of our family members as well as their binary best-fits. We also note that we characterized SIMP J0136+09334 as a member of this family, and this brown dwarf has been flagged as an overluminous variable. Recent retrieval fits to 1-15 $\mu$m spectral of SIMP J0136 found patchy silicate clouds in its atmosphere, consistent with the cloud-clearing theory \citep{Vos_2023,mcCarthy24}.

\section{Conclusion}

We have fit $\sim$300 archival spectra from the SpeX Prism Library with atmosphere models from the Sonora and Phoenix groups. We identified general trends and discussed four families of deviant objects. Here we summarize some of our main observations.

\begin{enumerate}
    \item Clouds have a more significant impact on near-infrared spectra than disequilibrium chemistry for L and T-type objects. 
    \item Our results agree with the literature that cloud clearing causes the effective temperature to plateau until a brown dwarf has evolved through the L/T transition.
    \item The same cloud clearing in conjunction with the condensation of methane causes a sudden switch to bluer colors past the L/T transition.
    \item Surface gravity appears to be poorly constrained by fits to the entire near-infrared spectrum.
    \item Early and mid L-type objects have thicker cloud decks made of smaller grains. The L/T transition brings about a transition in cloud properties, after which any clouds in T dwarf atmospheres are {thinner/deeper} in the atmosphere but have larger grains. 
    \item {Our fits suggest that the impact of silicate clouds on the near-infrared spectrum lingers through the late T-types, even though the 8-10 $\mu$m feature is not seen after L8} \citep{suarez22}.
    \item {We find that the Sonora models give a decreasing trend in both log($g$) and [M/H] with spectral type. This makes little physical sense, but could have to do with the degeneracy in the independent impacts of gravity and metallicity on CIA of hydrogen.}
    \item Analysis of disequilibrium chemistry through the L and T spectral types requires a suite of models including both clouds and multiple options for the extent of vertical mixing. Neither the Sonora models nor the suite of Phoenix models used in this project are sufficient for this aim. 
    \item The triangular \textit{H}-band profile indicative of low gravity tends to be linked to red \textit{J-K} colors. Young, low-gravity objects have thicker cloud decks.
    \item A plateaued \textit{H}-band profile could be indicative of binarity, but binarity alone does not completely explain the flat peak shape. Another parameter is likely at play here, such as significant vertical mixing causing disequilibrium carbon chemistry.
    \item A double-peaked \textit{H}-band profile is a likely indication of binarity. {Specifically, these binary systems likely include an L-type primary and a T-type secondary. The spectral type difference between the primary and secondary objects is greater for this family than for the plateaued H-band family.}
    \item Blue Early T-dwarfs likely are experiencing inhomogeneous cloud clearing, allowing light from deeper levels of the atmosphere to shine through cloud holes and mix with light at the surface. This mixing of light can cause these objects to be flagged as binaries.
\end{enumerate}

\subsection{Future Work}
In this era of the JWST and the influx of new spectral observations of exoplanets and brown dwarfs, model-fitting spectral surveys such as this will continue to be important tools in refining our understanding of the physical properties of planetary atmospheres and updating our modeling practices. There is more work to be done moving forward.

As more JWST mid-infrared brown dwarf spectra are released into the archives, we plan to extend our model fits to include longer wavelengths. This could help us to better disentangle the effects of surface gravity, disequilibrium chemistry, and clouds. 

We also plan to perform a more detailed investigation into the properties of the Double-Peak, Plateau, and Blue Early T families. We hope to develop spectral indices that can be used to find more members of each family in the archives. Fitting members from each family with flexible retrieval models could also help us to fine-tune their best-fit parameters and better pinpoint what makes these objects interesting. 

Future studies could also extend this survey into the M and Y spectral types, allowing us to evaluate model fits and parameter trends through the entire brown dwarf evolutionary sequence.

\section{Acknowledgments}
{We thank the anonymous referee for their thoughtful insight and helpful suggestions on how to improve this work.}

{We thank Caroline Morley for providing us with the beta version of the Diamondback models used in this work, and also for her help interpreting the trend in $f_{sed}$ discussed in Section \ref{subsec:clouds}.}

S.T. acknowledges funding from the Utah NASA Space Grant Consortium, NASA Grant 80NSSC20M0103.

This research has benefited from the SpeX Prism Spectral Libraries, maintained by Adam Burgasser at \url{http://pono.ucsd.edu/$\sim$adam/browndwarfs/spexprism}.

This work has made use of data from the European Space Agency (ESA) mission {\it Gaia} (\url{https://www.cosmos.esa.int/gaia}), processed by the {\it Gaia} Data Processing and Analysis Consortium (DPAC, \url{https://www.cosmos.esa.int/web/gaia/dpac/consortium}). Funding for the DPAC has been provided by national institutions, in particular the institutions participating in the {\it Gaia} Multilateral Agreement.

This research has made use of the SIMBAD database, operated at CDS, Strasbourg, France.

This project also benefited from the SIMPLE database at \url{https://simple-bd-archive.org/}.

\startlongtable
\begin{deluxetable}{ccccccccccc}
\tabletypesize{\footnotesize}
\label{tab:speXData}

\tablecaption{Archival Data}
\tablehead{\colhead{Object} & \colhead{OType} & \colhead{IRType} & \colhead{2MASS \textit{J}} & \colhead{2MASS \textit{H}} & \colhead{2MASS \textit{Ks}} & \colhead{Parallax} & \colhead{Sources} \\ 
\colhead{} & \colhead{} & \colhead{} & \colhead{} & \colhead{} & \colhead{} & \colhead{} & \colhead{} } 

\startdata
SDSS J000013.54+255418.6  &  T5  &  T4.5  & 15.063 & 14.731 & 14.836 & 70.8 & 66,9,9,26 \\
SDSS J000250.98+245413.8  &  ----  &  L5.5  & 17.165 & 16.06 & 15.656 &  ----  & 18,18 \\
LHS 102B (J00043484-4044058)  &  L5  &  L4.5  & 13.109 & 12.055 & 11.396 & 82.346602 & 45,50,17,35 \\
2MASS J00150206+2959323  &  L7  &  L7.5pec (blue)  & 16.158 & 15.226 & 14.482 & 31.6 & 47,47,47,4 \\
2MASS J00165953-4056541  &  L3.5  &  L4.5:  & 15.32 & 14.206 & 13.432 & 28.823598 & 46,13,13,35 \\
2MASS J00193927-3724392  &  L3:  &  L2.2 INT-G  & 15.52 & 14.47 & 13.689 & 25.008017 & 13,33,13,35 \\
2MASS J0028208+224905  &  ----  & L6 & 15.607 & 14.468 & 13.781 & 42.8879 & 82,13,35 \\
2MASSW J0030300-145033  &  L7  &  L6:FLD-G  & 16.278 & 15.273 & 14.481 & 41.1 & 44,55,13,4 \\
SDSSp J003259.36+141036.6  &  ----  &  L9  & 16.83 & 15.648 & 14.946 & 30.14 & 62,13,29 \\
\enddata

\tablecomments{List of all 301 archival spectra used in this project, along with their optical and infrared spectral types, 2MASS magnitudes, and parallaxes. Table \ref{tab:speXData} is published in its entirety in the machine-readable format.
      A portion is shown here for guidance regarding its form and content.}

\tablerefs{(1) \cite{Albert2011}; (2) \cite{Allers2013}; (3) \cite{Bard2014}; (4) \cite{Best2021}; (5) \cite{Best2021}; (6) \cite{BurgMcel2006}; (7) \cite{Burg03_1086}; (8) \cite{Burg03_510}; (9) \cite{Burg06}; (10) \cite{Burg07}; (11) \cite{Burg08}; (12) \cite{Burg09}; (13) \cite{Burg10}; (14) \cite{Burg04}; (15) \cite{Burg08b}; (16) \cite{Burg2011}; (17) \cite{burg07b}; (18) \cite{Chiu06}; (19) \cite{cruz04}; (20) \cite{cruz03}; (21) \cite{cruz07}; (22) \cite{Cruz_2009}; (23) \cite{Cruz18}; (24) \cite{cush05}; (25) \cite{dahn02}; (26) \cite{dupuy12}; (27) \cite{dupuy17}; (28) \cite{fahe10}; (29) \cite{fahe12}; (30) \cite{fahe13}; (31) \cite{Faherty_2016}; (32) \cite{fan00}; (33) \cite{gagn15}; (34) \cite{gaia2description}, \cite{gaiaDR2}; (35) \cite{2021}; (36) \cite{gauz19}; (37) \cite{Geballe_2002}; (38) \cite{gel10}; (39) \cite{gizi00}; (40) \cite{gizis01}; (41) \cite{gizi02}; (42) \cite{hawley02}; (43) \cite{kend04}; (44) \cite{kirk00}; (45) \cite{kirk01}; (46) \cite{Kirkpatrick_2008}; (47) \cite{kirk10}; (48) \cite{Kirkpatrick_1999}; (49) \cite{Kirkpatrick_2006}; (50) \cite{knap04}; (51) \cite{lieb03}; (52) \cite{lieb06}; (53) \cite{lieb07}; (54) \cite{Liu13}; (55) \cite{liu16}; (56) \cite{looper07}; (57) \cite{looper08}; (58) \cite{looperetal08}; (59) \cite{luhman570}; (60) \cite{manj19}; (61) \cite{maro13}; (62) \cite{maro15}; (63) \cite{mcel06}; (64) \cite{metch08}; (65) \cite{phan08}; (66) \cite{pine16}; (67) \cite{reid06}; (68) \cite{reid08}; (69) \cite{schmidt10}; (70) \cite{schn2014}; (71) \cite{sheppard09}; (72) \cite{siegler07}; (73) \cite{smart13}; (74) \cite{smart18}; (75) \cite{tinney03}; (76) \cite{tinney14}; (77) \cite{vrba04}; (78) \cite{wilson01}; (79) \cite{wilson03}; (80) \cite{zhan09}; (81) \cite{stumpf09}; (82) \cite{bard19}; (83) \cite{scholz02}; (84) \cite{schmidtb}}

\end{deluxetable}

\bibliography{sample631}{}
\bibliographystyle{aasjournal}

\end{document}